\documentclass[5p,twocolumn]{elsarticle}
\usepackage{graphicx,latexsym}
\usepackage{dcolumn}
\usepackage{amssymb,amsmath,bm}
\usepackage{subfigure}
\usepackage{braket}
\usepackage{siunitx}
\usepackage{array}
\usepackage{physics}
\usepackage{float}
\usepackage[nopar]{lipsum}
\usepackage{caption}
\usepackage{multirow}
\usepackage{bigstrut}

\biboptions{sort&compress}

\usepackage[super]{nth}

\usepackage{hyperref}
\hypersetup{
    pdfnewwindow=true,       % links in new window
    colorlinks=true,         % false: boxed links; true: colored links
    linkcolor=blue,          % color of internal links
    citecolor=blue,          % color of links to bibliography
    filecolor=magenta,       % color of file links
    urlcolor=black           % color of external links
}

\usepackage[normalem]{ulem}

\def\sec#1{Sec.\ \ref{#1}}
\def\eq#1{Eq.\ (\ref{#1})}
\def\fig#1{Fig.\ \ref{#1}}

\journal{}

\begin{document}

\begin{frontmatter}

%-----------------------------------------------------------------

\title{Thermal transport driven by Coulomb interactions in quantum dots: Enhancement of thermoelectric and heat currents}
	
\author[a1]{Bashdar Rahman Pirot}
\address[a1]{Physics Department, College of Education, 
	University of Sulaimani, Sulaimani 46001, Kurdistan Region, Iraq}

\author[a2,a3,a4]{Nzar Rauf Abdullah}
\ead{nzar.r.abdullah@gmail.com}
\cortext[correspondingauthor]{Corresponding author: Nzar Rauf Abdullah \\
	Postal address: Physics Department, College of Science, University of Sulaimani, Sulaimani 46001, Kurdistan Region, Iraq \\ Tel.: +964 770 144 3854.}
\address[a2]{Division of Computational Nanoscience, Physics Department, College of Science, 
	University of Sulaimani, Sulaimani 46001, Iraq}
\address[a3]{Computer Engineering Department, College of Engineering, Komar University of Science and Technology, Sulaimani, Iraq}
\address[a4]{Science Institute, University of Iceland, Dunhaga 3, IS-107 Reykjavik, Iceland}
\author[a1]{Ari Karim Ahmed}

%----------------------------------------------------------------

\begin{abstract}
	
We investigate thermal transport in a serial asymmetric double quantum dot (DQD)	coupled to two electron reservoirs with different temperatures. The inter- and intra-Coulomb interactions are taken into account in a Coulomb blockade DQD where the electron sequential tunneling via four different master equation approaches is considered.
In the absence of Coulomb interactions, a neglectable thermoelectric and heat currents is found  identifying as the Coulomb blockade DQD regime.
In the presence of Coulomb interactions, intra- and inter-Coulomb interactions, crossings energies between the intra- and the inter-dot many-body electron states are observed.
The crossings induce extra channels in the energy spectrum of the DQD that enhance thermoelectric and heat currents. The extra channels form several peaks in the thermoelectric and heat currents in which intensity and position of the peaks depend on strength of the inter- and intra-dot Coulomb interactions. In addition, the problem of coherences and incoherences are studied using 
different approaches to the master equation, which are the first order von-Neumann, the Redfield, a first order Lindblad, and the Pauli methods.
We find that all methods give almost similar thermal transport when the role of the coherences is irrelevant in the DQD.

\end{abstract}


\end{frontmatter}

\section{Introduction} 

The interest of thermoelectrics has long been too inefficient to be costeffective in most applications because of the lack of high-performance materials \cite{6436699, Mao2021}. In the mid of 1990s, it had been theoretically shown that the thermoelectric efficiency could be increased through nanostructural systems \cite{PhysRevB.47.12727}. It thus motivates experimentalist
to display the proof-of-principle and high-efficiency materials \cite{doi:10.1179/095066003225010182}, and the thermoelectric properties of a quantum dot (QD) or  DQD have been the subject of a renewed interest in the last two decades \cite{Snyder2008, Talbo2017, BAGHERITAGANI2012765}. 
 
Thermoelectric transport of DQD has lately been a hot topic because of it's delta-shape density of states that could arise the “best thermoelectric material" and a potential thermoelectricity utilization \cite{doi:10.1021/nn2007817}. There are several interesting thermoelectric phenomena in the DQD systems such as the oscillations of thermal conductance caused by Coulomb interactions between electrons which leads to Coulomb blockade phenomena arising nonlinear thermoelectric conductance \cite{Staring_1993, Zimbovskaya_2016}. The Pauli spin blockade in DQD has been seen when the electron spin is taken into account which will play an important role in the electron occupation and thermal transport  \cite{PhysRevB.72.165308}. Furthermore, the electron-phonon interactions and the interplay between them may have a significant impact on thermoelectric transport, and 
it results in phonon-assisted transport, enabling the conversion of local heat into electrical power in a nanosized heat engine \cite{doi:10.1021/acs.nanolett.0c04017}.

The coupled nano-systems such as double quantum dots in series \cite{PhysRevB.72.205319, PhysRevB.80.165333} and double quantum wires \cite{PhysicaE.64.254, PhysRevB.82.195325} have gained increasing attention due to their mutual electron-electron interactions \cite{PhysRevB.82.085311}. In addition to the Coulomb interaction in DQD, the density of state discreteness of DQD provides a fine control of current arising the Coulomb blockade effect that reduces thermal losses, and the reduction of DQD sizes down to a few nanometers can improve it's physical properties and operate it at room-temperature displaying a great Coulomb oscillations \cite{Mao2016, doi:10.1063/1.3483618, SVILANS20161096}. The quantum interference and Coulomb correlation effects in spin-polarized transport through two coupled quantum dot have been investigated \cite{PhysRevB.76.165432}, and the charge localization and isospin blockade in vertical double quantum dots are seen \cite{PhysRevB.70.081314}.
Therefore, one may ask an interesting question that follows from the aforementioned developments is how coupling and correlated carrier properties influences the thermoelectric and heat flow? or  
can such nano-systems give a good thermoelectric behavior?.

To answer these questions, we consider serially-coupled asymmetric DQD that are connected to two metallic leads when different types of Coulomb interactions are considered in the system. 
In this work, we demonstrate the influences of inter- and intra-dot Coulomb interactions on thermoelectric and heat currents of an asymmetric DQD. We show a great enhancement of thermal transport due to present both types of Coulomb interactions in the DQD. In addition, four different approaches to master equation are considered to investigate thermal transport of the system \cite{PIROT2022413646, AHMED2022413607}.

The current work is arranged as follows: in \sec{section_model} the Hamiltonian of DQD and electron reservoirs, and the master equations formalism are demonstrated. In \sec{section_results} the main obtained results under Coulomb interactions are presented. In \sec{section_conclusion}, the  conclusion is shown.

\section{Hamiltonian and Electron evolution Formalism}\label{section_model}

Our model is a DQD which is weakly connected to two metallic leads with different thermal gradient.

\subsection{System Hamiltonian}

The Hamiltonian of the total system, the DQD and the leads, is given by \cite{Goldozian2016, Abdullah2017}

\begin{equation}
\hat{H}_{\rm Total} = \hat{H}_{\rm DQD} + \hat{H}_l + \hat{H}_{\rm T}
\label{eq_Htotal}
\end{equation}

where $\hat{H}_{\rm DQD}$ refers to the Hamiltonian of the DQD, $\hat{H}_l$ displays the Hamiltonian of the $l$ lead, and $\hat{H}_{\rm T}$ indicates the tunneling Hamiltonian between the DQD and the leads. All terms of \eq{eq_Htotal} are further described below. 
It is assumed that there is two asymmetric double quantum dots in which 
the energy levels of the left dot is $E_{\rm L}$ and $E_{\rm L} + \Delta E_{\rm L}$, and the energy of the right lead is $E_{\rm R}$.
The Hamiltonian of DQD can be presented as \cite{doi:10.1021/acs.nanolett.0c04017}.

\begin{equation}
	\hat{H}_{\rm DQD} = \hat{H}_{\rm L} + \hat{H}_{\rm R} +  \hat{H}_{\rm \Omega}
	\label{H_DQD}
\end{equation}

where $\hat{H}_{\rm L}(\hat{H}_{\rm R})$ is the Hamiltonian of the left(right) quantum dot, and 
$\hat{H}_{\rm \Omega}$ defines the coupling between the quantum dots.
One can define the left quantum dot Hamiltonian in occupation number representation as

\begin{align}
\hat{H}_{\rm L} &= \sum_{\sigma}  E_{\rm L} \, \hat{a}^{\dagger}_{Lg\sigma} \hat{a}_{Lg\sigma}+( E_{\rm L} + \Delta E_{\rm L} ) \, \hat{a}^{\dagger}_{Le\sigma} \hat{a}_{Le\sigma}
                \nonumber \\
	            & + \frac{V_{\rm m}}{2} \sum_{i i^{\prime} j j^{\prime}} \sum_{\sigma \sigma^{\prime}}
	            \hat{a}^{\dagger}_{Li\sigma} \hat{a}^{\dagger}_{Li^{\prime}\sigma^{\prime}}
	            \hat{a}_{Lj^{\prime}\sigma^{\prime}} \hat{a}_{Lj\sigma}.
	            \label{H_L}
\end{align}

Herein, $\hat{a}^{\dagger}_{Li\sigma}$($\hat{a}_{Li\sigma}$) indicates the spin-dependent, $\sigma$, creation(annihilation) operator in the left dot, $i = e,g$, $\Delta E_{\rm L} = 0.005$~meV, and $i^{\prime}$ is introduced by $g^{\prime}=e$, $e^{\prime}=g$, and $V_{\rm m}$ is a constant intradot Coulomb interaction. The Hamiltonian of the right quantum dot is treated analogously with $L \rightarrow R$ except that there is only one energy level in the right dot, $E_{\rm R}$. This implies that we have two asymmetric double quantum dots.
The last term of \eq{H_DQD} is the coupling Hamiltonian between the two dots which can be defined by

\begin{align}
\hat{H}_{\Omega} & = \sum_{ii^{\prime}, \sigma} \Omega \, \hat{a}^{\dagger}_{Ri^{\prime}\sigma} \hat{a}_{Li\sigma} + {\rm h.c.} \nonumber \\
                 & + \sum_{ii^{\prime}, \sigma\sigma^{\prime}} V_{\rm n} \,
                  \hat{a}^{\dagger}_{Li{\sigma}}  \hat{a}^{\dagger}_{Ri^{\prime}{\sigma^{\prime}}}
                  \hat{a}_{Ri^{\prime}{\sigma^{\prime}}}  \hat{a}_{Li{\sigma}}.
                  \label{H_sigma}
\end{align}

where $\Omega$ indicates the inter-dot tunnel coupling strength, and $V_{\rm n}$ demonstrates the inter-dot Coulomb interaction between the quantum dots. Both the intra- and inter-Coulomb matrix elements are treated as the same way presented in \cite{Goldozian2016}.

The Hamiltonian of the metallic leads can be defined as 

\begin{equation}
\hat{H}_l = \sum_{q \sigma l} \varepsilon_{q \sigma l} \, d^{\dagger}_{q \sigma l} \, d_{q \sigma l} 
\end{equation}

$\varepsilon_{q \sigma l}$ is the spin-dependent energy levels of the leads, $d^{\dagger}_{q \sigma l}$($d_{q \sigma l}$) are the electron creation(annihilation) operators in the leads with index $l$, and $q$ indicates the spatial wave-functions of the continuum of states.

The DQD and the leads are connected via tunneling Hamiltonian, $\hat{H}_{\rm T}$, which is defined by

\begin{equation}
\hat{H}_{\rm T} = \sum_{i,q \sigma l} t_{il} \, a^{\dagger}_{i\sigma} d_{q \sigma l} + {\rm h.c.}
\end{equation}

where $t_{il}$ express the tunneling amplitude  \cite{KIRSANSKAS2017317}.

\subsection{Electron evolution formalism}

The electron evolution of the total system is described by density operator ($\rho(t)$)~\cite{GUDMUNDSSON20181672, 0953-8984-30-14-145303}, and the electron evolution 
of our system obeys the quantum Liouville-von Neumann equation

\begin{equation}
	\frac{\partial \rho(t)}{\partial t} = -i [\hat{H}_{\rm Total}, \rho(t)].
\end{equation}

In principle, the first order von Neumann, 1vN, covers only sequential tunneling in the case of coherences. In the 1vN, one of the conditions is that the coupling strength between the DQD and leads has to be smaller than the temperature of the leads $\Gamma_{L,R} << T_{L,R}$ \cite{Goldozian2016}.
One more condition of 1vN is that it can violate the positivity of $\rho_\mathrm{S}$ \cite{Goldhaber-Gordon1998}. 

We are interested in the dynamic of electrons of DQD, the reduced density operator, $\rho_{\rm S}(t)$, defining the electrons in the DQD under the influence of both leads is taken into account  \cite{Haake1973,Breuer2002}. The $\rho_{\rm S}(t)$ can be calculated by tracing out the variables of the leads \cite{Nzar_2016_JPCM, ABDULLAH2018}
\begin{equation}\label{master_equation}
	\rho_\mathrm{S}(t) = Tr_\mathrm{leads} \left\{ \rho (t) \right\}.
\end{equation}

The QmeQ package \cite{KIRSANSKAS2017317} can be used to solve \eq{master_equation} which gives $\rho_\mathrm{S}(t)$ in the interacting many-body Fock basis of the DQD, and it is assumed that $\hbar = 1.0$ and $\kappa_B = 1.0$ in the Qmeq software.

Once the $\rho_\mathrm{S}(t)$ is obtained, one can calculate the thermoelectric properties of the DQD in the steady state regime under applying a thermal gradient via a temperature difference between the leads \cite{PhysRevB.90.115313-2}.
The thermoelectric current, TEC (I$^{\rm TEC}$), through the lead channel $j$ is given by
\begin{equation}
	I^{\rm TEC}_{j} =  -\frac{\partial}{\partial t} \expval{N_{j}}
	= -i \expval{[\hat{H}_{\rm Total}, N_{j}]},
	\label{I_TEC}
\end{equation}
with $N_{j} = \sum_q d^{\dagger}_{j q} d_{j q}$. 
The energy current, EC ($I^{\rm EC}_{j}$), is introduced using
\begin{equation}
	I^{\rm EC}_{j}  = -\frac{\partial}{\partial t} \expval{H_j}
	= -i\expval{[\hat{H}_{\rm Total}, H_{j}]},
	\label{I_HC}
\end{equation}
where $H_{j} = \sum_q \varepsilon_{jq} \, d^{\dagger}_{j q} d_{j q}$.

\section{Results}\label{section_results}

The main obtained results of energy spectrum, occupation, and thermal transport of the DQD are presented in this section. The temperature gradient or the thermal energy, $k_BT_{\rm L}(k_BT_{\rm R})$, of the leads are assumed to be $1.5$($0.5$)~meV. 
We also consider a small interdot tunnel coupling strength $\Omega = 0.05$~meV, and the coupling strength between the DQD and the leads is $\Gamma_{\rm L,R} = 90*10^{-6}$~meV. The condition of $\Gamma_{L,R} << T_{L,R}$ is thus considered in our calculations \cite{Kouwenhoven_2001}, which is one of the condition of 1vN. The 1vN is implemented in QmeQ in which $k_B = 1.0$ is assumed. 
The thermal gradient applied to the leads causes different energy distributions of the leads around the electrochemical potentials $\mu$ in which the chemical potential of both leads are considered to be equal to $\mu_L = \mu_R = 0.0$.
 
\subsection{DQD with no Coulomb interactions}
 
\begin{figure}[htb]
	\centering
	\includegraphics[width=0.45\textwidth]{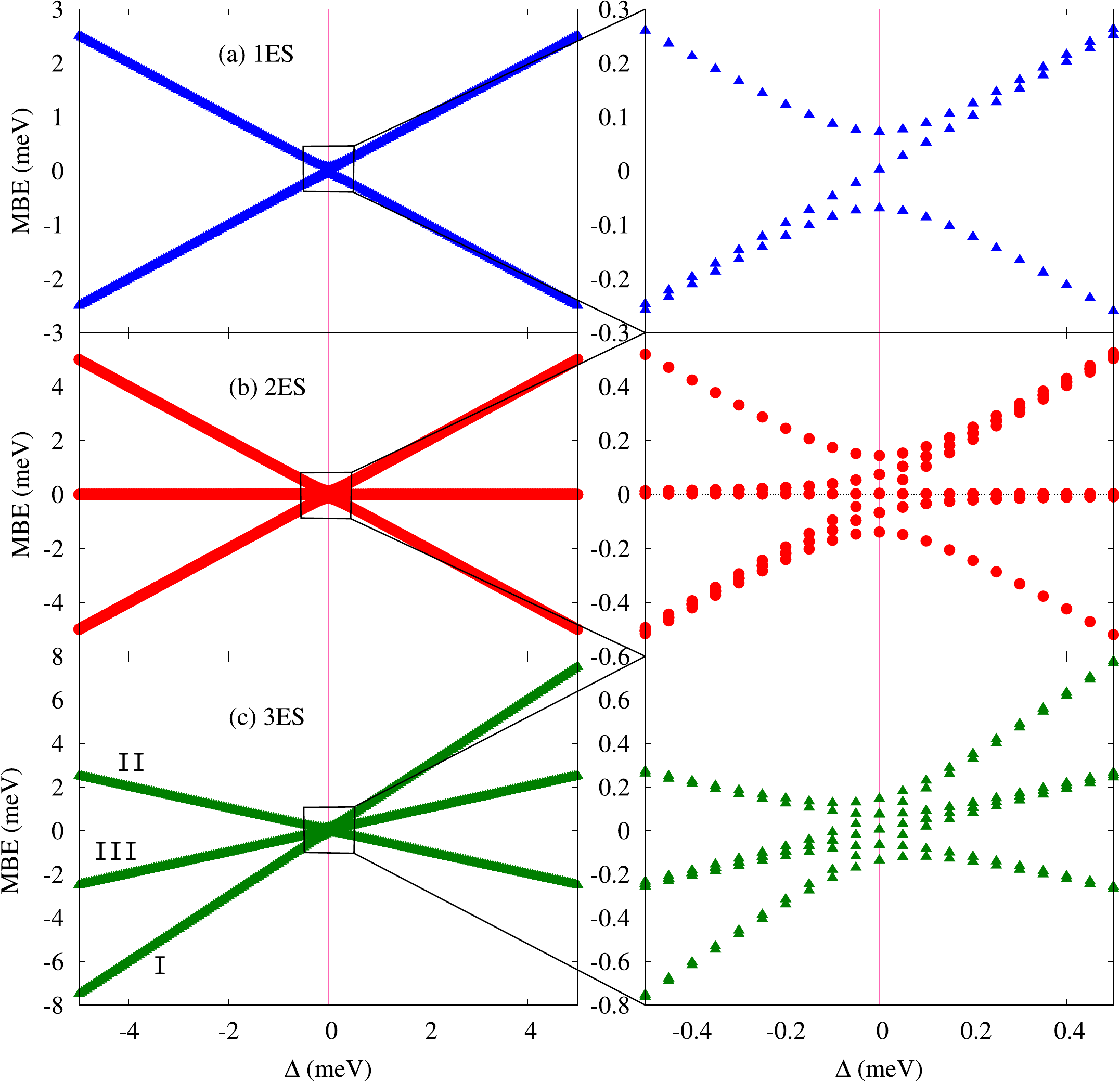}  
	\caption{Many-Body Energy (MBE) spectrum versus $\Delta = E_{\rm L}\text{-}E_{\rm R}$ for (a) one-electron state, 1ES, (blue triangle), (b) two-electron states, 2ES (red circles), and (c) three-electron states, 3ES, (green diamond) for the asymmetric DQD without Coulomb interactions, V$_{\rm col} = 0.0$. The vertical pink line is the position when $E_{\rm L} = E_{\rm R}$. 
	The right panel is just the MBE spectrum in the smaller range of $\Delta$. 
	There are two types of 2ES: They can have both electrons in the same dot (two red circles that are changed with $\Delta$) or one electron in either dot (the red circles that are unchanged with $\Delta$.}
	\label{fig01}
\end{figure}

We first assume the Coulomb interactions shown in \eq{H_L} and \eq{H_sigma} are neglected in the DQD, $V_m = 0.0$ and $V_n = 0.0$. The many-body energy (MBE) of the DQD as a function of $\Delta$ is presented in \fig{fig01} for 1ES (a), 2ES (b), and 3ES (c), where $\Delta=E_{\rm L} \text{-} E_{\rm R}$. In order to see the energy states more clear, the \fig{fig01} is re-plotted on the smaller scale of $x$- and $y$-axis in the right panel.
One can clear see an energy crossing in all three types of energy states at $\Delta = 0.0$ (vertical pink line) indicating the resonance energy between the left, $E_{\rm L}$, and the right, $E_{\rm R}$, quantum dots, $E_{\rm L} \approx E_{\rm R}$.

At $\Delta < 0.0$, the number of 1ES states with negative value of energy below chemical potential, $\mu = 0.0$~meV, (dashed horizontal line at zero axis) are twice of the 1ES with positive value located above the chemical potential. This confirms that the energy states of the left quantum dots are twice of the energy states of the right dot, as the states below $\mu$ shows energy levels of the left dot at $\Delta < 0.0$. 
In contrast, at positive value of $\Delta$ ($\Delta > 0.0$), the energy states below $\mu$ indicate the energy of the right dot, while the energy states above $\mu$ display the energy levels of the left dot. 
We should mentioned that  each 1ES, and 3ES is double degenerate due to the spin of the electrons, $\sigma = \, \uparrow, \downarrow$. 

There are two types of 2ES, states with one electron in each dot have a flat dispersion in Fig.\ \ref{fig01}(b) identifying as inter-dot 2ES, and the states with both electrons in the same dot identifying as intra-dot 2ES have a strong dispersion. 
In \fig{fig01}(c), the crossing of 3ES at $\Delta=0.0$ is also seen in addition to an asymmetric  distribution of energy states around zero value of $x$-axis or the $\mu$. 
There are three types of 3ES: 
First, the states with energy of $\approx \text{-}7.5$~meV at $\Delta = \text{-}5.0$~meV extend to $7.5$~meV at $\Delta = 5.0$~meV, identifying as 3ES-I. Second, the states have energy value of $\approx2.5$~meV at $\Delta = \text{-}5.0$~meV extending to $\approx\text{-}2.5$~meV at $\Delta = 5.0$~meV, identifying as 3ES-II.
These two types of 3ES contains four states including spin-up and down.  
Third, the 3ES have energy value of $\approx\text{-}2.5$~meV at $\Delta = \text{-}5.0$~meV extending to $\approx2.5$~meV at $\Delta = 5.0$~meV, which contains twelve 3ES identifying as 3ES-III.

\begin{figure}[htb]
	\centering
	\includegraphics[width=0.45\textwidth]{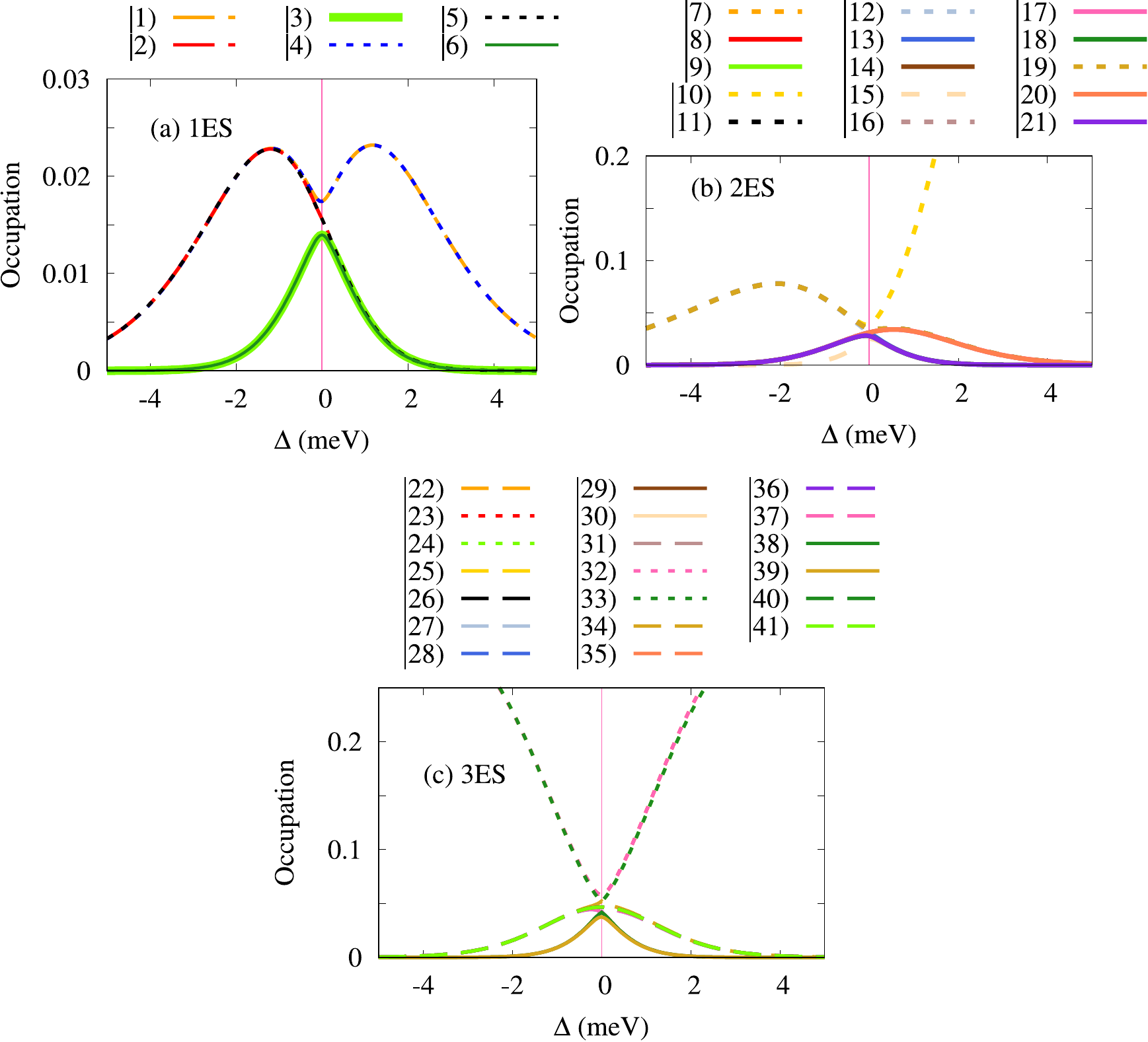} 
	\caption{Partial occupation as a function of $\Delta$ for the one-electron state, 1ES, (a), the two-electron states, 2ES (b), and the three-electron states, 3ES, (c) of the DQD without Coulomb interactions, V$_{\rm col} = 0.0$. The chemical potentials of the leads are $\mu_{\rm L} = \mu_{\rm R} = 0.0$, which coincides with the vertical pink line when $\Delta = E_{\rm L}=E_{\rm R} = 0.0$.In Fig. (b), dashed lines are the occupation of the 2ES when both electrons are found in the same dot, and solid lines are the occupation of the 2ES when one electron is in either dot. The thermal energy of the leads are assumed to be $k_BT_L = 1.5$~meV and $k_BT_R = 0.5$~meV, and the coupling strength is $\Gamma_{L,R} = 90\times10^{-6}$~meV.}
	\label{fig02}
\end{figure}

The occupation or partial occupation of DQD with no Coulomb interaction is presented in \fig{fig02} for 1ES (a), 2ES (b), and 3ES (c). We have found the occupation of six 1ES (three spin-down and three spin-up) in \fig{fig02}(a) including states from $|1)$ to $|6)$. It can be clearly seen that the states $|1)$, $|2)$, $|4)$, and $|5)$ are the energy states of the left quantum dot at $\Delta < 0.0$, as they are occupied and their occupation in increased from $\Delta \approx -5$ to $-1.5$~meV. The states of $|3)$ and $|6)$ are the 1ES of the right quantum dot at $\Delta < 0.0$ in which they are unoccupied from $\Delta \approx -5$ to $-1.5$~meV. This is expected as there are four 1ES below $\mu$ and two 1ES above $\mu$ shown in \fig{fig01}(a). The 1ES below $\mu$ are occupied due to thermal energies of the left leads. By further tuning the value of $\Delta$, we found that the 1ES of the left quantum dot are depopulated and the 1ES of the right dot are populated at $\Delta = 0.0$ which caused by the energy crossings of 1ES of both dots at this point.
At the positive value of $\Delta > 0$, the states of $|1)$ and $|4)$ become the 1ES of the right which are occupied at lower value of positive $\Delta$, and other states are the 1ES of the left dot which are thus depopulated. 

The occupation of 2ES shown in \fig{fig02}(b) are classified as follows: The occupation of inter-dot 2ESs (solid and dashed lines) and the occupation of intra-dot 2ES (dotted lines). The intra-dot 2ES located below $\mu$ are occupied and the occupation is increased from $\Delta = -5.0$ to $-2.0$~meV, while the intra-dot 2ES above $\mu$ is unoccupied in the same range of $\Delta$. In addition, the inter-dot 2ES with a flat dispersion are unoccupied from $\Delta = -5.0$ to $-2.0$~meV. 
Interestingly, the intra-dot 2ES are depopulated and the inter-dot 2ES are populated at $\Delta = 0.0$~meV confirming a strong resonance or crossing of these two types of 2ESs.  

The occupation of 3ES presented in \fig{fig02}(c) are also classified according to their energy positions with respect to the chemical potential of the leads, $\mu$: The 3ES-I are occupied (dotted lines) while the 3ES-II (solid lines) and 3ES-III (dashed lines) are not occupied at $\Delta = -5.0$~meV. The occupation of 3ES-I is decreased with increasing $\Delta$, and the occupation of 
3ES-II and 3ES-III are increased until they are crossing at $\Delta = 0.0$~meV.
The dashed lines indicating the occupation of the 3ES-III seem to play similar role of inter-dot 2ES.

\begin{figure}[htb]
	\centering
	\includegraphics[width=0.45\textwidth]{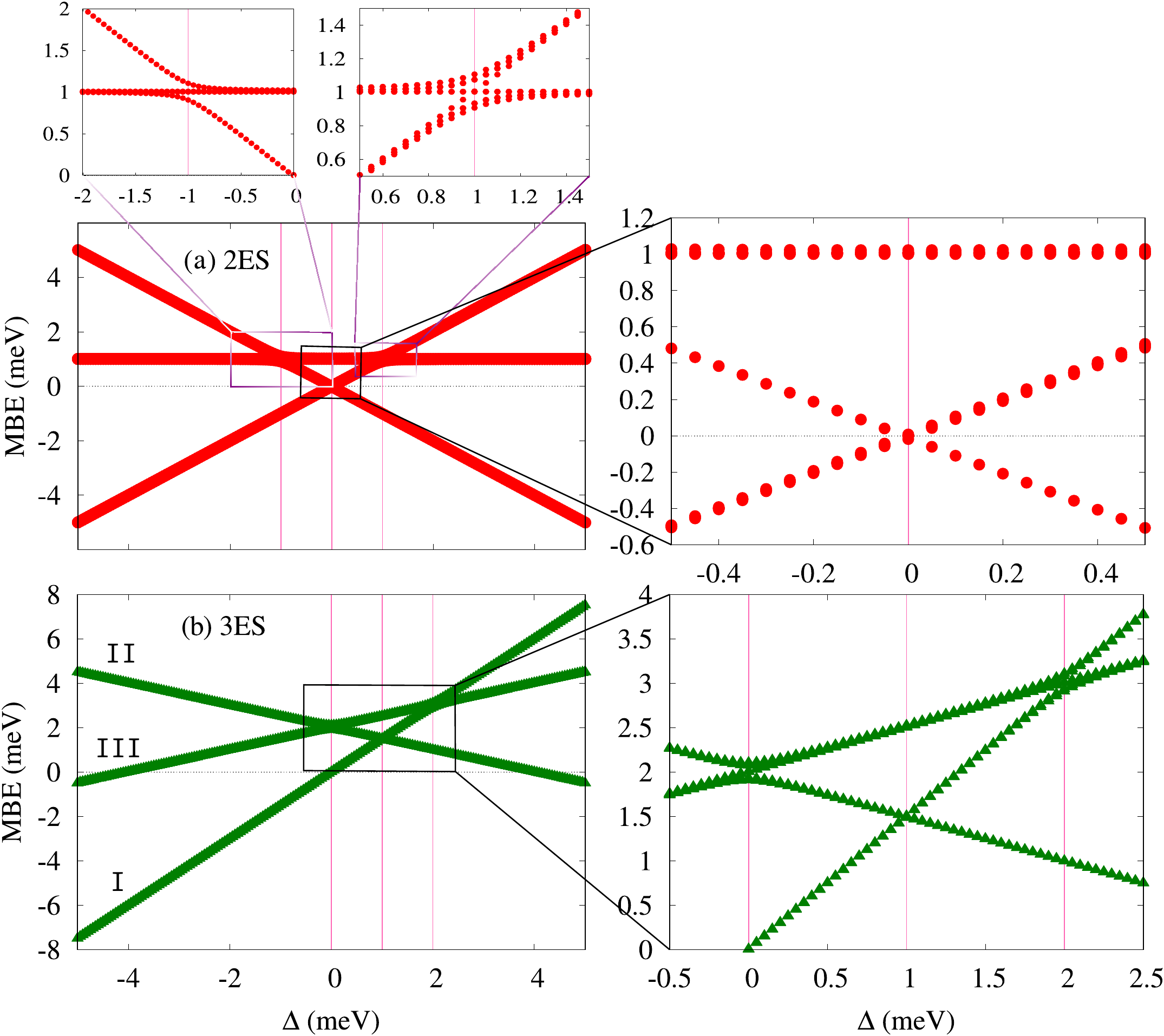}  %MBE_Uintra_0_Uinter_0
	\caption{Many-Body Energy (MBE) spectrum versus $\Delta$ for (a) two-electron states, 2ES (red circles), and (b) three-electron states, 3ES, (green diamond) for the asymmetric DQD in the presence of only the inter-dot Coulomb interaction, V$_{\rm n} = 1.0$~meV. The vertical pink lines are the resonance or crossing energies positions. 
		The right panel is just the MBE spectrum in the smaller range of $\Delta$. 
		The 2ESs representing one electron in either dot are shifted up by the value of V$_{\rm n} = 1.0$~meV, while the 2ESs representing the electrons in the same dot are not changed.}
	\label{fig03}
\end{figure}

\subsection{DQD with inter-dot Coulomb interaction}

We now consider only the inter-dot Coulomb interaction, $V_n$, in the DQD and neglect the intra-dot Coulomb interaction, $V_m$, and we assume the strength of inter-dot Coulomb interaction to be $V_n = 1.0$~meV. 
Figure \ref{fig03} indicates the MBE for the 2ES (a), and 3ES (b) in the presence of inter-dot Coulomb interaction. The right panel is nothing but it is just the energy spectrum of the left panel on the smaller scale. It should be first mentioned that the 1ES is not affected by the inter-dot Coulomb interaction and it's energy spectrum is the same as of the 1ES spectrum shown in \fig{fig01}(a).

It can be clearly seen that the inter-dot 2ESs with a flat dispersion are now shifted up by $V_n = 1.0$~meV forming the energy crossings between inter- and intra-dot 2ES at $\Delta = \text{+}V_n$ (most-right vertical pink line) in addition to the energy crossings of only intra-dot 2ES at $\Delta = 0.0$. A better view of the energy crossings at $\Delta = \text{+}V_n$ is shown in the right inset of the right panel of \fig{fig03}(a). We should mention that the inter- and intra-dot 2ES at $\Delta = \text{-}V_{n}$  (most-left vertical pink line) are closing to each other but they are not crossing (see left inset of the right panel).

\begin{figure}[htb]
	\centering
	\includegraphics[width=0.45\textwidth]{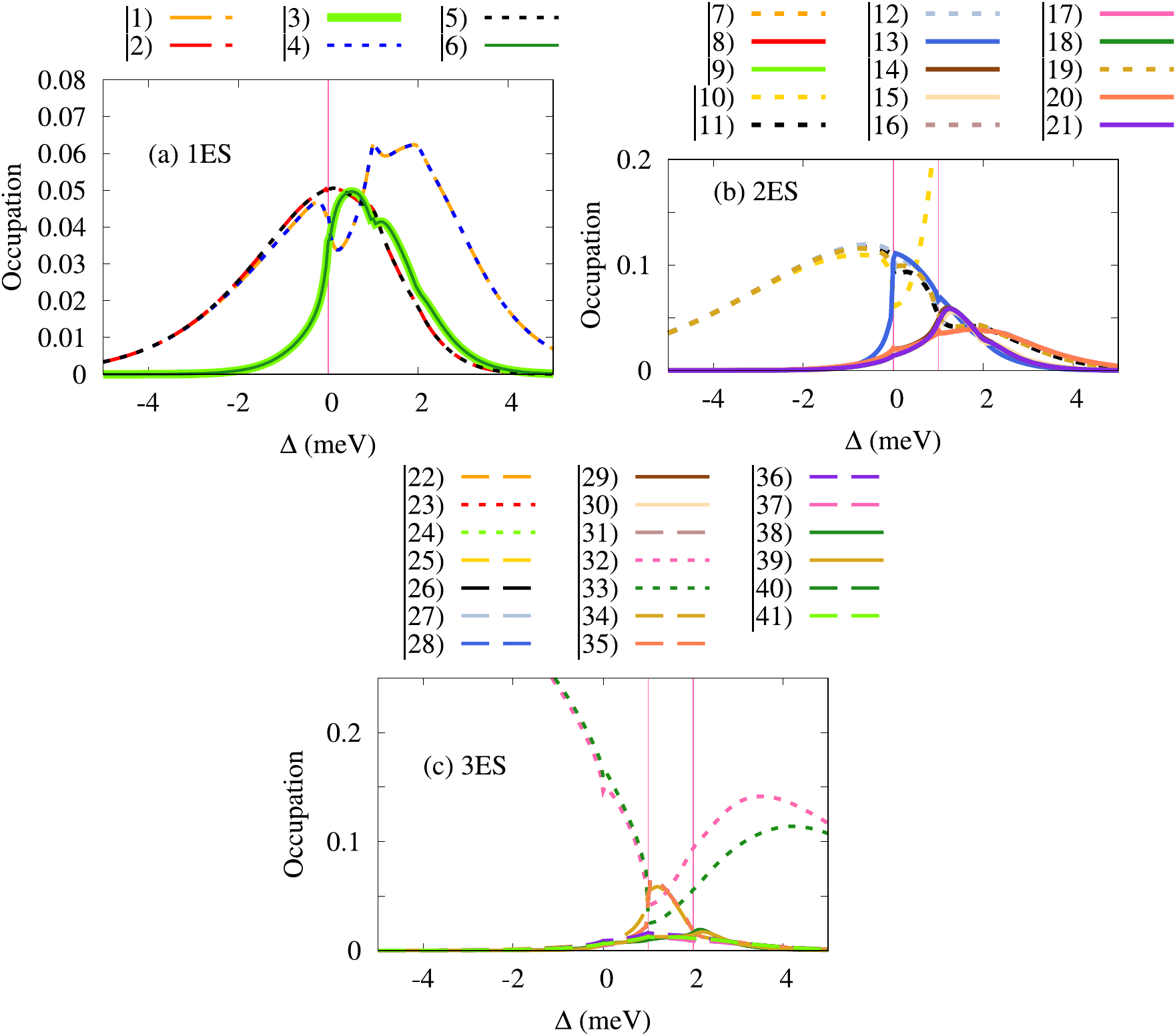}  %MBE_Uintra_0_Uinter_0
	\caption{Partial occupation as a function of $\Delta$ for the one-electron state, 1ES, (a), the two-electron states, 2ES (b), and the three-electron states, 3ES, (c) of the DQD in the presence of only inter-dot Coulomb interactions, V$_{\rm n} = 1.0$~meV. The chemical potentials of the leads are $\mu_{\rm L} = \mu_{\rm R} = 0.0$, which coincides with the vertical pink line when $\Delta = E_{\rm L}=E_{\rm R} = 0.0$.
		In Fig. (b), dashed lines are the occupation of the 2ES when both electrons are found in the same dot, and solid lines are the occupation of the 2ES when one electron is in either dot. The thermal energy of the leads are assumed to be $k_BT_L = 1.5$~meV and $k_BT_R = 0.5$~meV, and 
		the coupling strength is $\Gamma_{L,R} = 90\times10^{-6}$~meV.}
	\label{fig04}
\end{figure}  

The inter-dot Coulomb interaction influences the 3ES-II and the 3ES-III, and they are shifted up resulting  energy crossing with 3ES-I at $\Delta = \text{+}V_n$, and $V_n\text{+}1$ as they are shown in \fig{fig03}(b).
We should mention that the 3ES-III are not affected by inter-dot Coulomb interaction. The energy spectrum of 2ES and 3ES in the presence of inter-dot Coulomb interactions indicate that the energy crossings are shifted to the positive axis of $\Delta$, which means that the energy states of the left quantum dot must be higher than that of the right dot. 

The occupation of 1ES (a), 2ES (b), and 3ES (c) is demonstrated in \fig{fig04} when the inter-dot Coulomb interaction is considered in the DQD. As we just mentioned that the 1ES spectrum is not influenced by $V_n$, and the occupation of 1ES is thus not much affected by the inter-dot Coulomb interaction in which the energy resonances is only seen at $\Delta = 0.0$ with a very low occupation (see \fig{fig04}(a)). 
In contrast, the occupation of inter-dot and intra-dot 2ESs shown in \fig{fig04}(b) are crossing at $\Delta = 0.0$, and $+V_n$ with a high occupation. It is interesting to see no crossing in occupation at $\Delta = \text{-}V_n$ as there is no energy crossings of 2ES at $\Delta = \text{-}V_n$. At this point, the energy of the left dot is lower than that of the right dot resulting a Coulomb blockade of DQD. This is only appearing in an asymmetric DQD and this should not be occurred when a symmetric DQD is considered.

Furthermore, the energy crossings of 3ES shown in \fig{fig03}(b) induce a crossing of occupation at $\Delta = + V_n$, and $V_n + 1$ with a high occupation. We should mentioned that these extra channels formed due to inter-dot Coulomb interaction will influence thermal transport which will be demonstrated later. 

\begin{figure}[htb]
	\centering
	\includegraphics[width=0.45\textwidth]{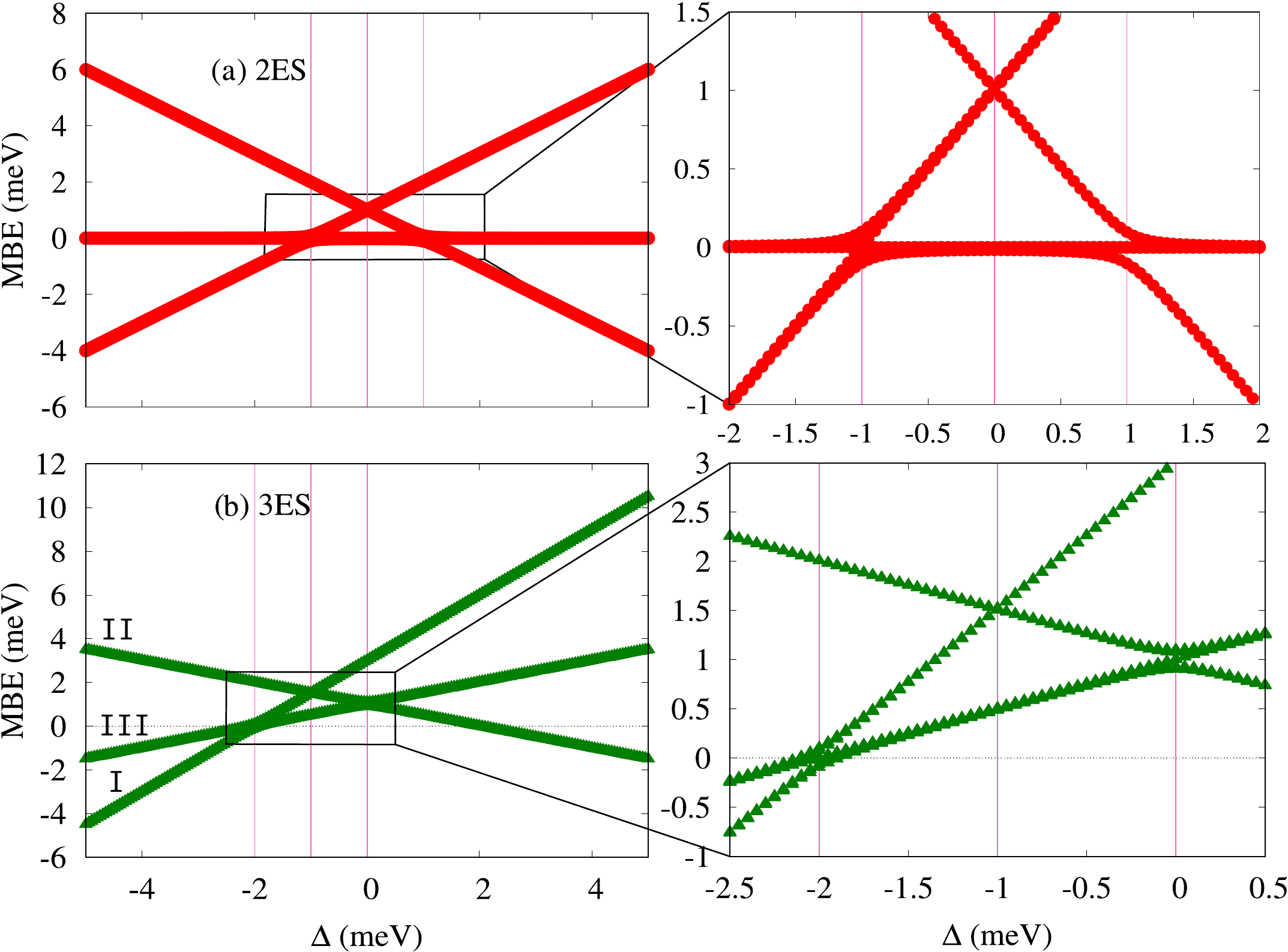}  %MBE_Uintra_0_Uinter_0
	\caption{Many-Body Energy (MBE) spectrum versus $\Delta$ for (a) two-electron states, 2ES (red circles), and (b) three-electron states, 3ES, (green diamond) for the asymmetric DQD in the presence of only the intra-dot Coulomb interaction, V$_{\rm m} = 1.0$~meV. The vertical pink lines are the resonance or crossing energies positions. 
		The right panel is just the MBE spectrum in the smaller range of $\Delta$. 
		The 2ESs representing one electron in either dot are unchanged with $\Delta$, while the 2ESs representing the electrons in the same dot are shifted up by the value of V$_{\rm m} = 1.0$~meV.}
	\label{fig05}
\end{figure}

\subsection{DQD with intra-dot Coulomb interaction}

We now consider the DQD with only intra-dot Coulomb interaction and neglecting the inter-dot Coulomb interaction, and assuming the intra-dot Coulomb interaction is $V_m = 1.0$~meV. The MBE of the DQD in the presence of intra-dot Coulomb interaction is shown in \fig{fig05} for 2ES (a), and 3ES (b). 
Similar to the DQD with the inter-dot Coulomb interaction, the energy spectrum of 1ES is not changed by the intra-dot Coulomb interaction.
In contrast to the presence of inter-dot Coulomb interaction,
the intra-dot 2ES are shifted up by $V_{m} = 1.0$~meV forming energy crossings between the intra-dot 2ES at $\Delta = 0.0$, and the intra- and inter-dot 2ESs at $\Delta = \text{-}V_{m}$. The inter- and intra-dot 2ES do not form energy crossings at $\Delta = +V_{m}$ which is totally opposite to the cases of inter-dot Coulomb interaction. 

\begin{figure}[htb]
	\centering
	\includegraphics[width=0.45\textwidth]{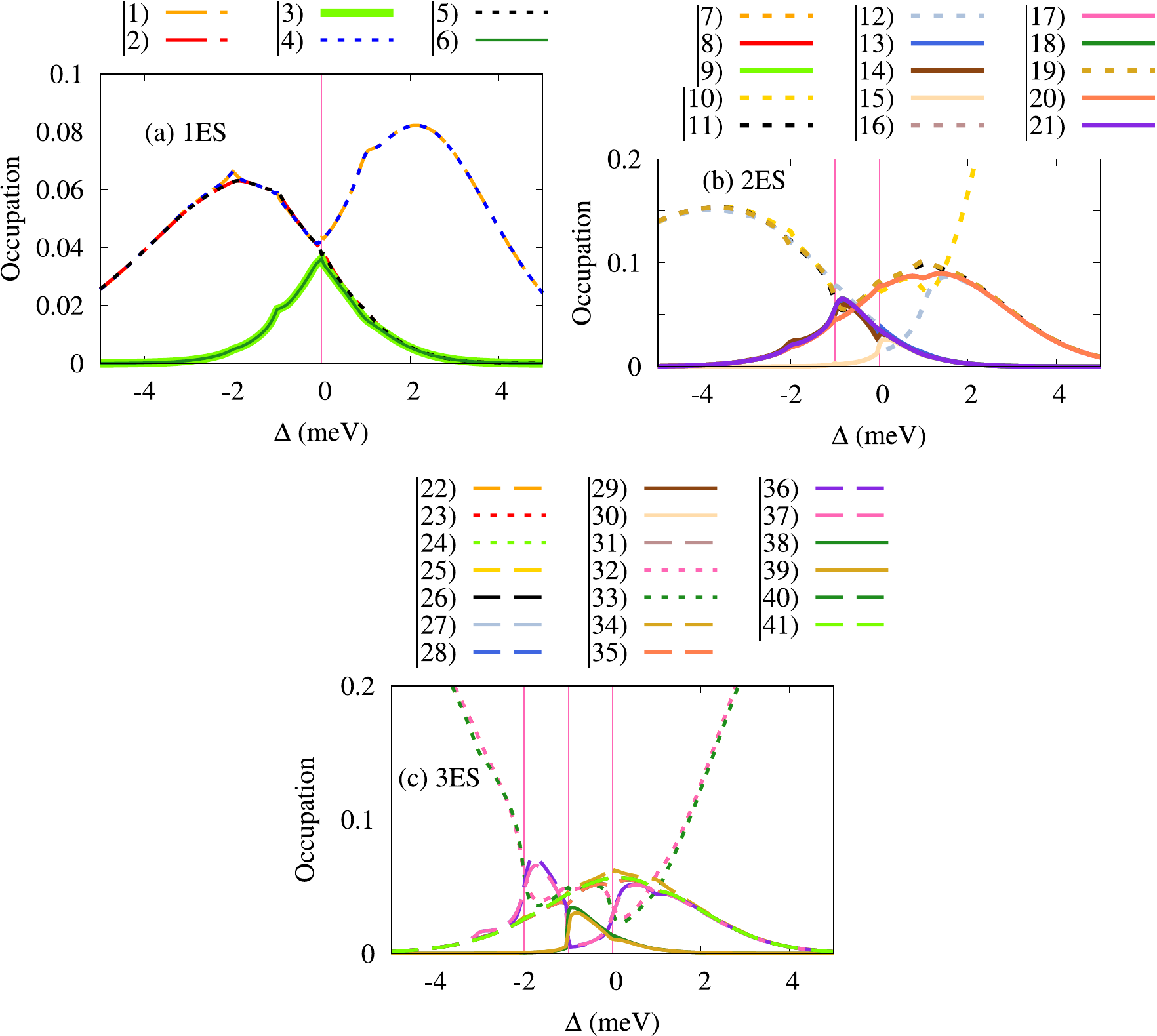}  %MBE_Uintra_0_Uinter_0
	\caption{Partial occupation as a function of $\Delta$ for the one-electron state, 1ES, (a), the two-electron states, 2ES (b), and the three-electron states, 3ES, (c) of the DQD in the presence of only intra-dot Coulomb interactions, V$_{\rm m} = 1.0$~meV. The chemical potentials of the leads are $\mu_{\rm L} = \mu_{\rm R} = 0.0$, which coincides with the vertical pink line when $\Delta = E_{\rm L}=E_{\rm R} = 0.0$.
		In Fig. (b), dashed lines are the occupation of the 2ES when both electrons are found in the same dot, and solid lines are the occupation of the 2ES when one electron is in either dot. The thermal energy of the leads are assumed to be $k_BT_L = 1.5$~meV and $k_BT_R = 0.5$~meV, and 
		the coupling strength is $\Gamma_{L,R} = 90\times10^{-6}$~meV.}
	\label{fig06}
\end{figure}

All three types of 3ES are shifted up in the presence of the intra-dot Coulomb interaction displayed in \fig{fig05}(b), but the energy shifting of 3ES-I is much stronger than that of 3ES-II and 3ES-III which is opposite to the 3ES in the presence of inter-dot Coulomb interaction shown in \fig{fig03}(b). 
The energy crossings of 3ES in the presence of inter-dot Coulomb interaction shown in \fig{fig03} were shifted to the positive axis of $\Delta$, while the same energy crossings of 3ES in the presence of intra-dot Coulomb interaction are seen in the negative axis of $\Delta$ at $\Delta=\text{-}V_{m}$ and $\text{-}V_{m}\text{-}1$ (see \fig{fig05}(b)). So, we can conclude that the energy crossings of the DQD in the presence of intra-dot Coulomb interaction is opposite to the inter-dot Coulomb interaction for both 2ES and 3ES spectrum.

Let's now look at the occupation of the DQD in the presence of intra-dot Coulomb interaction with strength of $V_{m} = 1.0$~meV which is displayed in \fig{fig06} for 1ES (a), 2ES (b), and 3ES (c). No big change in the occupation of 1ES is seen in the presence of $V_{m}$ which is expected as the energy spectrum of 1ES is not much influenced by intra-dot Coulomb interaction. 
The crossing in occupation between 1ES occur at $\Delta = 0.0$ with a low rate of occupation.

In contrast to the occupation of 2ES in presence of the inter-dot Coulomb interaction, population and depopulation of 2ES are seen at the negative axis of $\Delta$ at $\Delta = 0.0$, and $\text{-}V_{m}$. 
The occupation of 3ES shows that the occupation crossings are formed at $\Delta = \text{-}V_{m}\text{-}1$, $\text{-}V_{m}$, $0.0$, and $\text{+}V_{m}$ which has two more extra populated and depopulated positions comparing to the occupation of 3ES of the DQD in the presence of inter-dot Coulomb interaction.

\begin{figure}[htb]
	\centering
	\includegraphics[width=0.45\textwidth]{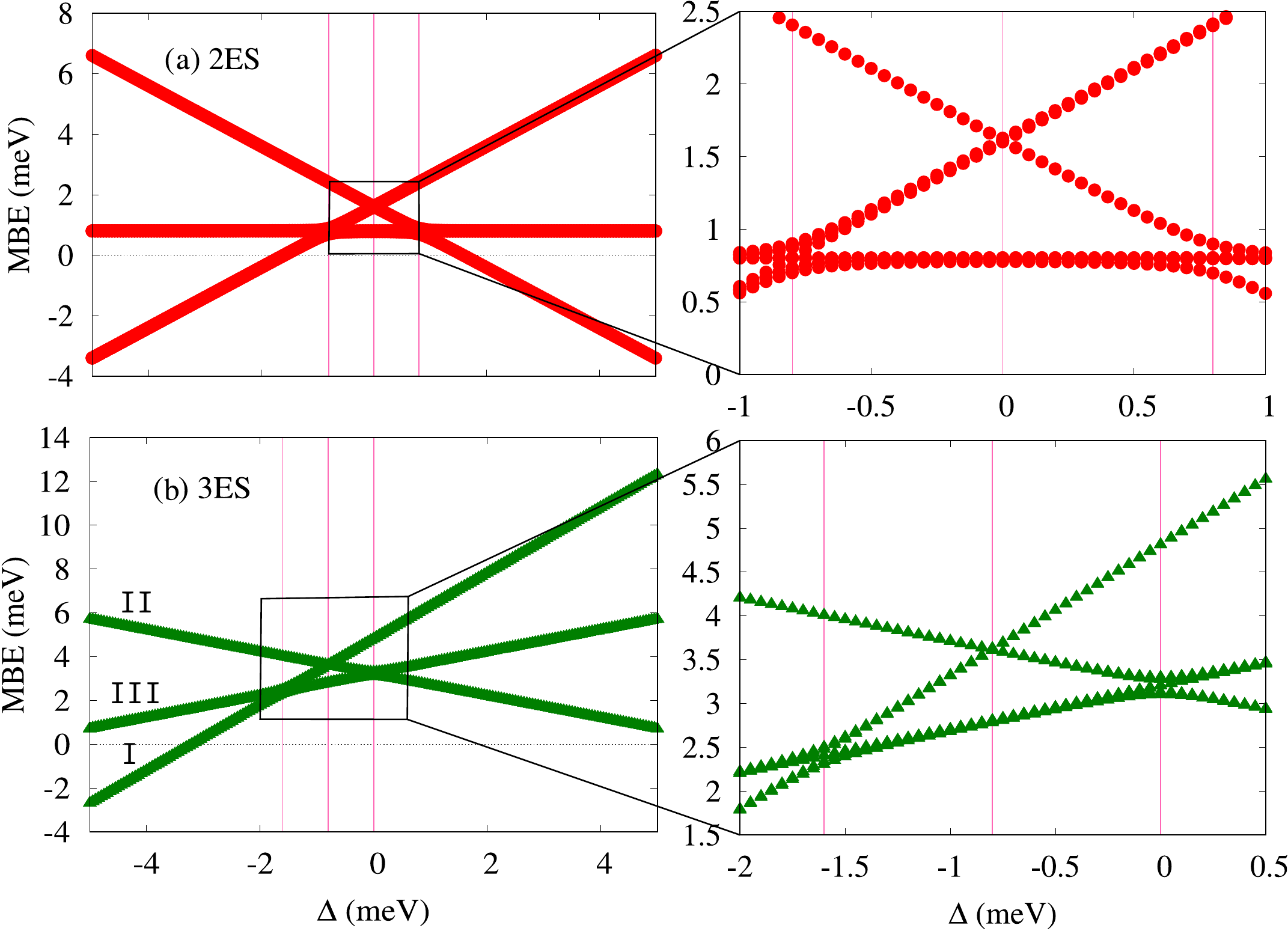}  %MBE_Uintra_0_Uinter_0
	\caption{Many-Body Energy (MBE) spectrum versus $\Delta$ for (a) two-electron states, 2ES (red circles), and (b) three-electron states, 3ES, (green diamond) for the asymmetric DQD in the presence of inter-dot, V$_{\rm n} = 0.8$~meV and the intra-dot Coulomb interaction, V$_{\rm m} = 1.6$~meV. The vertical pink lines are the resonance or crossing energies positions. 
		The right panel is just the MBE spectrum in the smaller range of $\Delta$.
		Both 2ESs are shifted up in which  the 2ESs representing one electron in either dot are shifted up by the value of V$_{\rm n} = 0.8$~meV, while the 2ESs representing the electrons in the same dot are shifted up by the value of V$_{\rm m} = 1.6$~meV.}
	\label{fig07}
\end{figure}

\subsection{DQD with both inter- and intra-dot Coulomb interactions}

In order to have more applicable DQD devices, we now consider both inter- and intra-dot Coulomb interaction in which the strength of inter-dot Coulomb interactions is $V_{n} = 0.8$~meV and the intra-dot Coulomb interaction is $V_{m} = 1.6$~meV. The MBE spectrum of the DQD in the presence of both types of Coulomb interaction is displayed in \fig{fig07} for 2ES (a), and 3ES (b).  
It can be clearly see that the inter-dot 2ES is shifted up by $V_n$ and the intra-dot 2ES are shifted up by $V_m$ forming energy crossings at $\Delta = \pm(V_m \text{-}V_n)$ in addition to the crossing at $\Delta = 0.0$. 
In the energy spectrum of 3ES, the effect of intra-dot Coulomb interaction is dominant because the strength of intra-dot Coulomb interaction is higher than that of the inter-dot Coulomb interaction. 
As a result, the energy crossings of 3ES are found at the negative value of $\Delta$.

\begin{figure}[htb]
	\centering
	\includegraphics[width=0.45\textwidth]{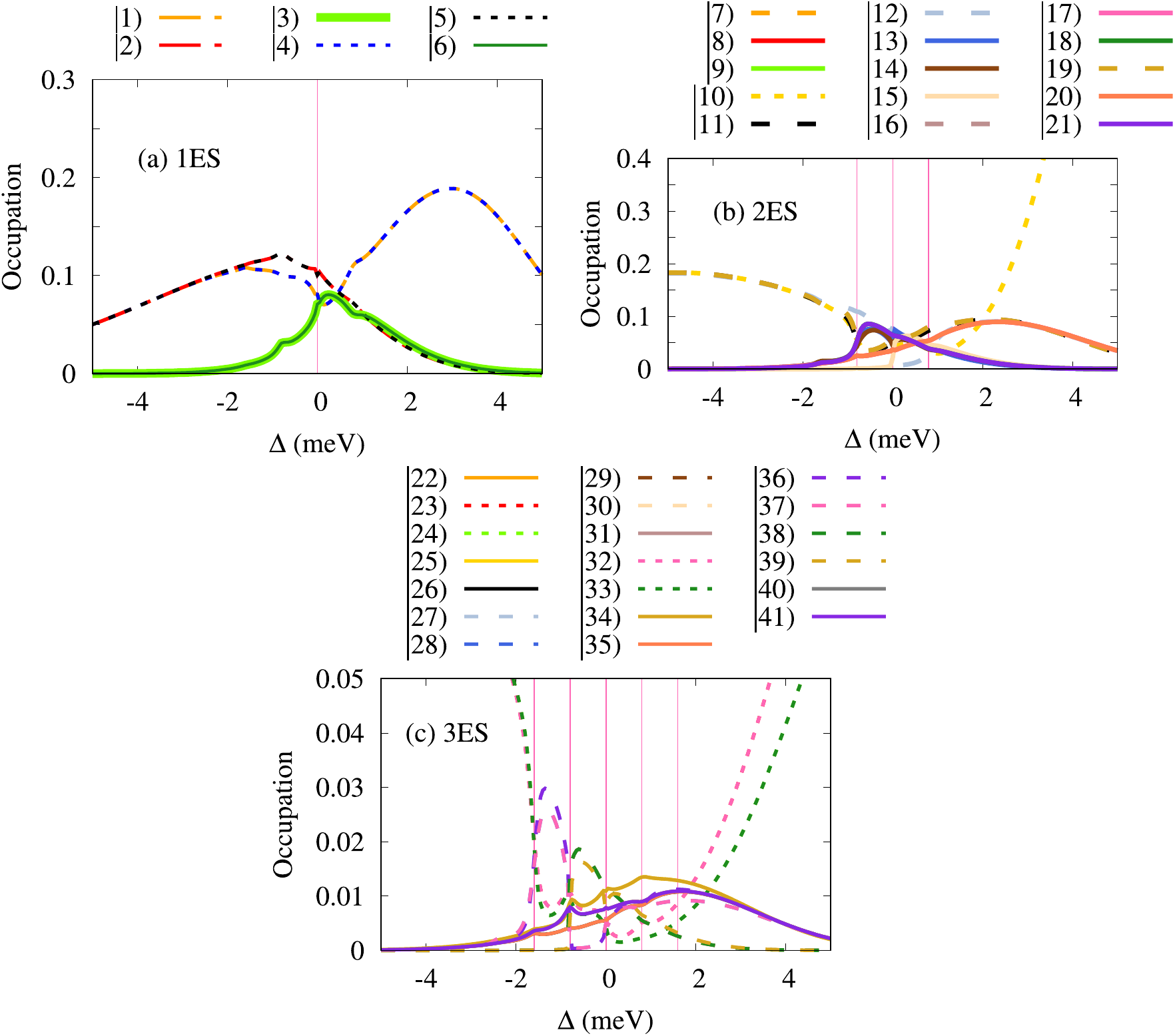}  %MBE_Uintra_0_Uinter_0
	\caption{Partial occupation as a function of $\Delta$ for the one-electron state, 1ES, (a), the two-electron states, 2ES (b), and the three-electron states, 3ES, (c) of the DQD in the presence of inter-dot, V$_{\rm m} = 0.8$~meV, and intra-dot Coulomb interactions, V$_{\rm m} = 1.6$~meV. The chemical potentials of the leads are $\mu_{\rm L} = \mu_{\rm R} = 0.0$, which coincides with the vertical pink line when $\Delta = E_{\rm L}=E_{\rm R} = 0.0$.
		In Fig. (b), dashed lines are the occupation of the 2ES when both electrons are found in the same dot, and solid lines are the occupation of the 2ES when one electron is in either dot. The thermal energy of the leads are assumed to be $k_BT_L = 1.5$~meV and $k_BT_R = 0.5$~meV, and 
		the coupling strength is $\Gamma_{L,R} = 90\times10^{-6}$~meV.}
	\label{fig08}
\end{figure}

The occupation of 1ES (a), 2ES (b), and 3ES (c) are demonstrated in \fig{fig08} for the DQD in the presence of both types of Coulomb interactions. The occupation of 1ES records a crossing at $\Delta = 0.0$ indicating the energy crossings between the left and the right quantum dots. The occupation of 1ES here is much higher than the occupation of 1ES shown in \fig{fig04} and \fig{fig06} which indicates that the 1ES are more effective here and contribute to the transport in the DQD.

The occupation of 2ES indicate three resonances positions between populated and depopulated intra- and inter-dot 2ES at $\Delta = 0$, and $\pm(V_m \text{-}V_n)$. All these three points in the energy spectrum of 2ES will effectively contribute to the transport as their occupations are high enough.
Furthermore, more positions of populated and depopulated 3ES are seen in the presence of both inter- and intra-dot Coulomb interactions comparing to the occupation of 3ES shown in \fig{fig04} and \fig{fig06}.

\subsection{Thermal transport}

\begin{figure}[htb]
	\centering
	\includegraphics[width=0.35\textwidth]{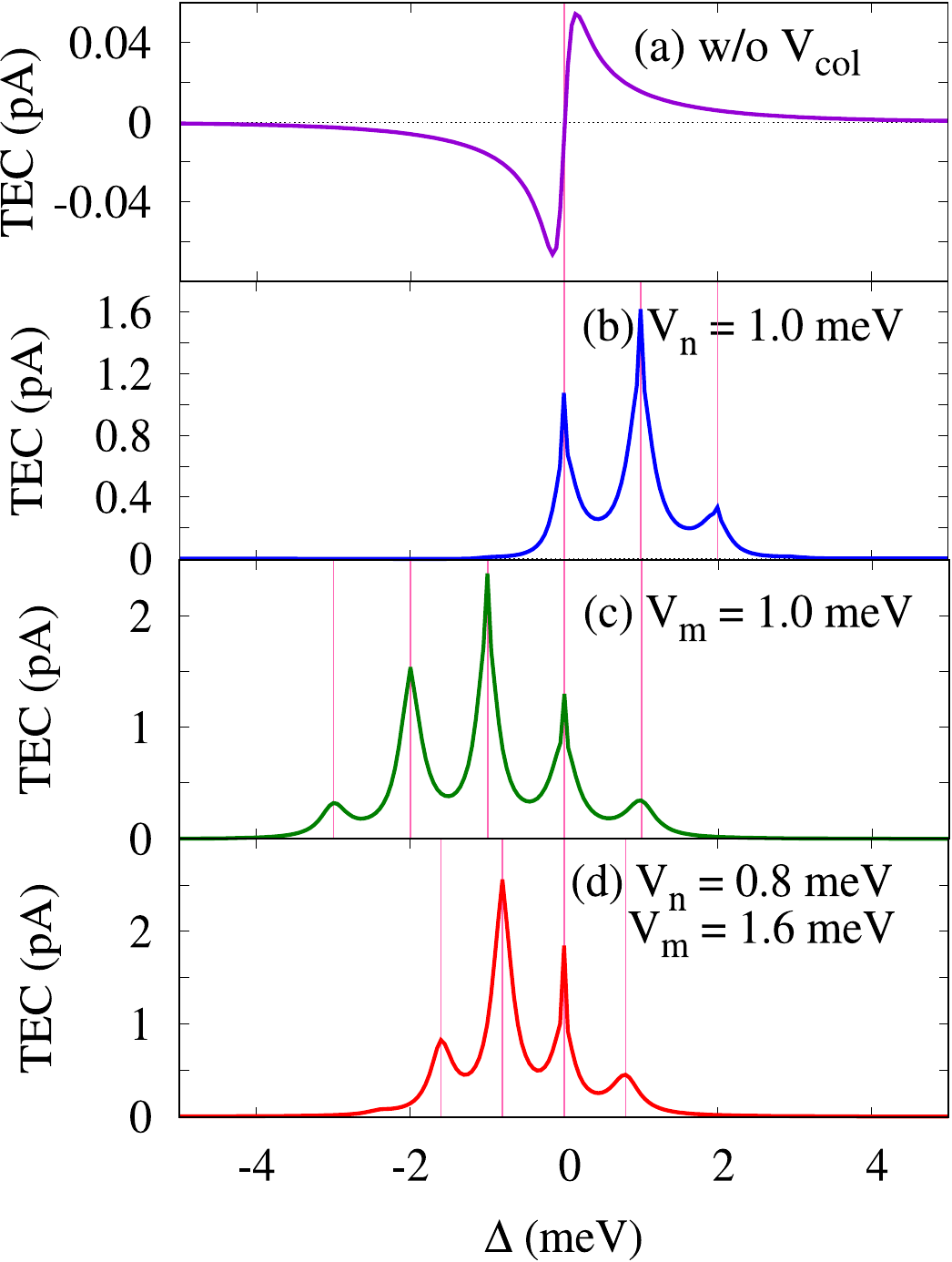} %MBE_Uintra_0_Uinter_0
	\caption{Thermoelectric current, TEC as a function of $\Delta$ for the DQD (a) without Coulomb interactions V$_{\rm col} = 0.0$~meV, (b) with only inter-dot Coulomb interactions of V$_{\rm n} = 1.0$~meV, (c) with only intra-dot Coulomb interactions of V$_{\rm m} = 1.0$~meV, and (d) with
	the inter-dot and intra-dot Coulomb interactions with strength of V$_{\rm n} = 0.8$ and V$_{\rm m} = 1.6$~meV, respectively.
	The chemical potentials of the leads are $\mu_L = \mu_R = 0.0$~meV which coincides with the middle vertical golden line when $\Delta = E_L = E_R = 0.0$.
	The thermal energy of the leads are assumed to be $k_BT_L = 1.5$~meV and $k_BT_R = 0.5$~meV, and the coupling strength is $\Gamma_{L,R} = 90 \times 10^{-6}$~meV.}
	\label{fig09}
\end{figure}

\begin{figure}[htb]
	\centering
	\includegraphics[width=0.35\textwidth]{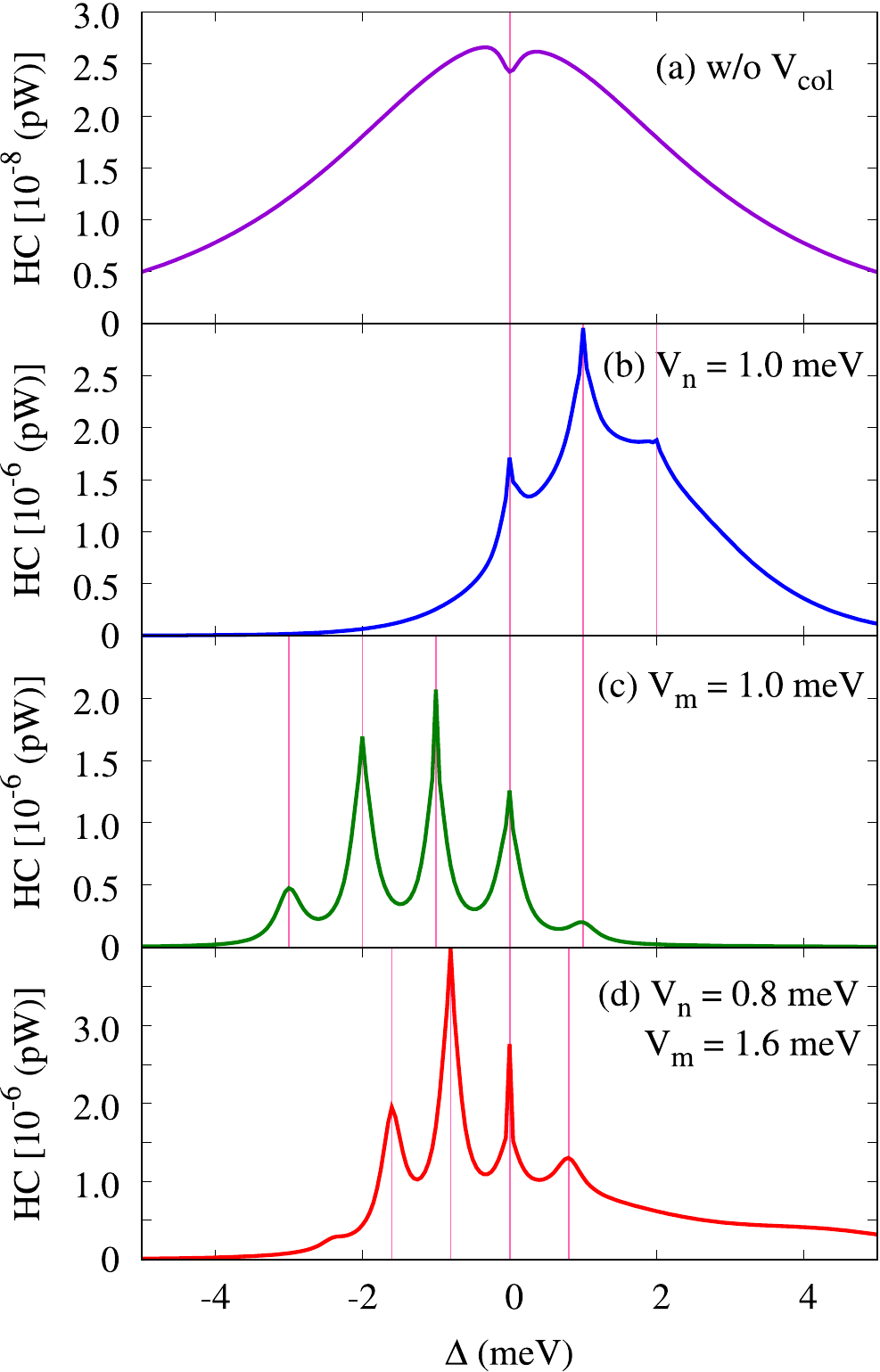}
	\caption{Heat current, HC as a function of $\Delta$ for the DQD (a) without Coulomb interactions V$_{\rm col} = 0.0$~meV, (b) with only inter-dot Coulomb interactions of V$_{\rm n} = 1.0$~meV, (c) with only intra-dot Coulomb interactions of V$_{\rm m} = 1.0$~meV, and (d) with
		the inter-dot and intra-dot Coulomb interactions with strength of V$_{\rm n} = 0.8$ and V$_{\rm m} = 1.6$~meV, respectively.
		The chemical potentials of the leads are $\mu_L = \mu_R = 0.0$~meV which coincides with the middle vertical golden line when $\Delta = E_L = E_R = 0.0$.
		The thermal energy of the leads are assumed to be $k_BT_L = 1.5$~meV and $k_BT_R = 0.5$~meV, and the coupling strength is $\Gamma_{L,R} = 90 \times 10^{-6}$~meV.}
	\label{fig10}
\end{figure}

In this section, we present the thermal transport properties of the DQD attached to two leads with 
different temperature mentioned above. The thermal transport such as thermoelectric current, TEC, (\fig{fig09}) and heat current, HC, (\fig{fig10}) are shown. The TEC ($I^{\rm TEC}$) and HC ($I^{\rm HC}$) are defined in \eq{I_TEC} and \eq{I_HC}, respectively.
In both \fig{fig09} and \fig{fig10}, the TEC and HC are displayed for the DQD in the absence of both types of Coulomb interactions, $V_{\rm col} = 0.0$ (a), and in the presence of only inter-dot Coulomb interaction with strength of $V_{n} = 1.0$~meV (b), only intra-dot Coulomb interaction with strength of $V_{m} = 1.0$~meV (c), and inter- and intra-dot Coulomb interaction with strength of $V_{n} = 0.8$~meV, and $V_{m} = 1.6$~meV (d), respectively. 

The TEC in the absence of Coulomb interaction has negative value at $\Delta < 0.0$, and positive value at $\Delta > 0.0$, and zero value of TEC is recorded at $\Delta=0.0$ when the chemical potential of the leads are $\mu = 0.0$. This is expected as the MBE states (1ES, 2ES, and 3ES), are asymmetrically distributed above and below $\mu$ in which the state below $\mu$ gives rise a negative thermoelectric current while the states above $\mu$ lead to generating positive current. 
In addition, of the chemical potential of the leads resonance with the states of the DQD, a zero-value of TEC has to be obtained as it is shown in \fig{fig09}(a). 
The heat current in the absence of the Coulomb interaction should be zero when the chemical potential of the leads are in resonance with states of the DQD, and it should have positive value around each state. We can see that the HC is very small (Atto Watt) in which the HC has positive value around the energy crossings and should approves to zero at $\Delta = 0.0$~meV \cite{Nzar_ACS2016}. But the zero value of HC in our system is not seen as there is a broad energy crossings at $\Delta = 0.0$~meV (see \fig{fig10}(a)) \cite{ABDULLAH2018199}.

In the presence of inter-dot Coulomb interaction with strength of $V_{n} = 1.0$~meV, three peaks in TEC and HC are found at $\Delta = 0.0$, $\text{+}V_n$, and $\text{+}(V_n+1)$. 
The first peak at $\Delta = 0.0$~meV is formed due to the contribution of 1ES and 2ES, and the second peak at $\Delta =\text{+}V_n$ is generated by 2ES and 3ES, while the third peak at $\Delta=\text{+}(V_n+1)$ is totally formed by 3ES. We can clearly see it from energy crossings and partial occupation shown in both \fig{fig03} and \fig{fig04}.

In the presence of only intra-dot Coulomb interaction in DQD with strength of $V_{m} = 1.0$~meV, five peaks in TEC (\fig{fig09}(c)) and  HC (\fig{fig10}(c)) are clearly found. From the energy spectrum and the occupation of MBE states in the presence of intra-dot Coulomb interaction shown early, we can distinguish and recognize the reasons for generating each peak in TEC and HC here. 
The peak at $\Delta = \text{-}V_{m}\text{-}2$ is formed due to four-electron states (not shown), the peak at $\Delta = \text{-}V_{m}\text{-}1$ refers to only 3ES, the peak at $\Delta = \text{-}V_{m}$ is raised by 2ES and 3ES, the peak at $\Delta = 0.0$ is generated by 1ES, 2ES, and 3ES, and finally the peak at $\Delta = \text{+}V_{m}$ refers to the 3ES. It is intersting to mention that the number of peak in both TEC and HC in the presence of intra-dot Coulomb interaction is higher than that of the inter-dot Coulomb interaction.

The TEC (\fig{fig09}(d)) and HC (\fig{fig10}(d)) in the presence of both the inter- and the intra-dot Coulomb interactions have four main peaks at $\Delta = \text{-}V_{m}$, $0.0$, and $\pm(V_m \text{-} V_n)$ where $V_n = 0.8$~meV, and $V_m = 1.6$~meV. From the partial occupation demonstrated in \fig{fig08}, we can see that the peak at $\Delta = \text{-}V_{m}$ refers to only 3ES, and the peaks at $\Delta = \pm(V_m \text{-} V_n)$ are formed due to 2ES and 3ES, and the peak at $\Delta = 0.0$ is formed by all three types of energy states, 1ES, 2ES, and 3ES.

\begin{figure}
	\centering
	\includegraphics[width=0.4\textwidth]{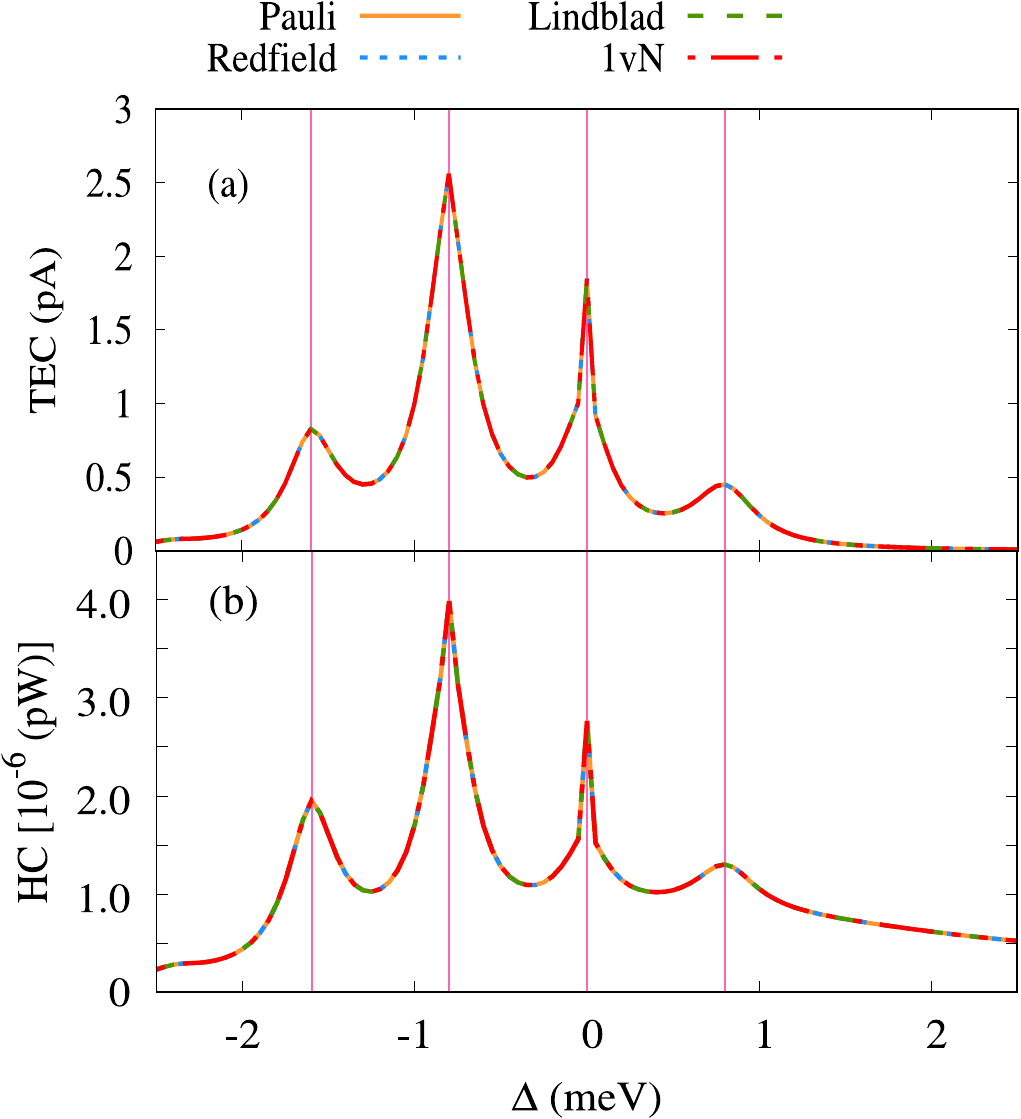} 
	\caption{Thermoelectric current (a), and heat current (b) as a function of $\Delta$ for different approches to master equation where the inter-dot and intra-dot Coulomb interactions are considered with strength of V$_{\rm n} = 0.8$ and V$_{\rm m} = 1.6$~meV, respectively.
	The chemical potentials of the leads are $\mu_L = \mu_R = 0.0$~meV which coincides with the middle vertical golden line when $\Delta = E_L = E_R = 0.0$.
	The thermal energy of the leads are assumed to be $k_BT_L = 1.5$~meV and $k_BT_R = 0.5$~meV, the coupling strength is $\Gamma_{L,R} = 90 \times 10^{-6}$~meV, and $\Omega = 0.05$~meV.}
	\label{fig11}
\end{figure}

Finally, we assume different sequential tunneling approaches of master equations 
in \fig{fig11} including the Pauli \cite{Grabert1982}, the Redfield \cite{PhysRev.89.728}, and the simple first order Lindblad approach \cite{PEDERSEN2010595, JONSSON201781} in addition to the 1vN method. The approximations in these sequential approaches are almost the same except that the sequential tunneling in the presence of coherences is taken into account for the Redfield, the first order Lindblad approach, and the 1vN approaches, while the coherences in the Pauli formulation are neglected. If the coherences are relevant in a DQD system, the appropriate master equation approach should be considered by taking care the values of $\Omega$ and $\Gamma$ as the coherences are mainly affected by these two parameters of the system \cite{KIRSANSKAS2017317}. 
If $\Gamma_{\rm L,R} < \Omega$, the coherences are irrelevant, while in the case of $\Gamma_{\rm L,R} > \Omega$, the role of the coherences is relevant. We can therefore see that a perfect agreement between the Pauli current and the current obtained via the other methods of master equations in \fig{fig11} when the coherences are irrelevant in our system.

\section{Conclusions}\label{section_conclusion}

We have studied the physical properties such as many-body energy spectrum, occupations or partial occupations, and thermoelectric and heat currents of an asymmetric double quantum dots. 
Several form of master equations implemented in the QmeQ software are used to investigate the DQD coupled to two matallic leads. We found that the inter- and intra-dot Coulomb interaction play an essential role in thermoelectric and heat currents of the DQD, and an influential enhancement in the thermal transport is recorded in the presence of Coulomb interactions. In addition, we could determine the most active one, two, and three many-body electron states to the transport. 
Our study may be benefit for thermoelectric devices of nanoscale range.

\section{Acknowledgment}
This work was financially supported by the University of Sulaimani and 
the Research center of Komar University of Science and Technology. 
BRP would like to thank Prof. Andreas Wacker from Lund University for his help and discussions, and NRA would like to thank Prof. Vidar Gudmundsson from Iceland University for his continuous fruitful discussions.
The computations were performed on resources provided by the Division of Computational 
Nanoscience at the University of Sulaimani.  
 
%\section{References}

%\bibliographystyle{elsarticle-num} 
%\bibliography{Ref.bib}

\begin{thebibliography}{10}
	\expandafter\ifx\csname url\endcsname\relax
	\def\url#1{\texttt{#1}}\fi
	\expandafter\ifx\csname urlprefix\endcsname\relax\def\urlprefix{URL }\fi
	\expandafter\ifx\csname href\endcsname\relax
	\def\href#1#2{#2} \def\path#1{#1}\fi
	
	\bibitem{6436699}
	W.~C. White, Some experiments with peltier effect, Electrical Engineering
	70~(7) (1951) 589--591.
	\newblock \href {https://doi.org/10.1109/EE.1951.6436699}
	{\path{doi:10.1109/EE.1951.6436699}}.
	
	\bibitem{Mao2021}
	J.~Mao, G.~Chen, Z.~Ren,
	\href{https://doi.org/10.1038/s41563-020-00852-w}{Thermoelectric cooling
		materials}, Nature Materials 20~(4) (2021) 454--461.
	\newblock \href {https://doi.org/10.1038/s41563-020-00852-w}
	{\path{doi:10.1038/s41563-020-00852-w}}.
	\newline\urlprefix\url{https://doi.org/10.1038/s41563-020-00852-w}
	
	\bibitem{PhysRevB.47.12727}
	L.~D. Hicks, M.~S. Dresselhaus,
	\href{http://link.aps.org/doi/10.1103/PhysRevB.47.12727}{{Effect of
			quantum-well structures on the thermoelectric figure of merit}}, Phys. Rev. B
	47 (1993) 12727--12731.
	\newblock \href {https://doi.org/10.1103/PhysRevB.47.12727}
	{\path{doi:10.1103/PhysRevB.47.12727}}.
	\newline\urlprefix\url{http://link.aps.org/doi/10.1103/PhysRevB.47.12727}
	
	\bibitem{doi:10.1179/095066003225010182}
	G.~Chen, M.~S. Dresselhaus, G.~Dresselhaus, J.-P. Fleurial, T.~Caillat,
	\href{https://doi.org/10.1179/095066003225010182}{Recent developments in
		thermoelectric materials}, International Materials Reviews 48~(1) (2003)
	45--66.
	\newblock \href
	{http://arxiv.org/abs/https://doi.org/10.1179/095066003225010182}
	{\path{arXiv:https://doi.org/10.1179/095066003225010182}}, \href
	{https://doi.org/10.1179/095066003225010182}
	{\path{doi:10.1179/095066003225010182}}.
	\newline\urlprefix\url{https://doi.org/10.1179/095066003225010182}
	
	\bibitem{Snyder2008}
	G.~J. Snyder, E.~S. Toberer, \href{https://doi.org/10.1038/nmat2090}{Complex
		thermoelectric materials}, Nature Materials 7~(2) (2008) 105--114.
	\newblock \href {https://doi.org/10.1038/nmat2090}
	{\path{doi:10.1038/nmat2090}}.
	\newline\urlprefix\url{https://doi.org/10.1038/nmat2090}
	
	\bibitem{Talbo2017}
	V.~Talbo, J.~Saint-Martin, S.~Retailleau, P.~Dollfus,
	\href{https://doi.org/10.1038/s41598-017-14009-4}{Non-linear effects and
		thermoelectric efficiency of quantum dot-based single-electron transistors},
	Scientific Reports 7~(1) (2017) 14783.
	\newblock \href {https://doi.org/10.1038/s41598-017-14009-4}
	{\path{doi:10.1038/s41598-017-14009-4}}.
	\newline\urlprefix\url{https://doi.org/10.1038/s41598-017-14009-4}
	
	\bibitem{BAGHERITAGANI2012765}
	M.~{Bagheri Tagani}, H.~{Rahimpour Soleimani},
	\href{https://www.sciencedirect.com/science/article/pii/S0921452611012142}{Thermoelectric
		effects through weakly coupled double quantum dots}, Physica B: Condensed
	Matter 407~(4) (2012) 765--769.
	\newblock \href {https://doi.org/https://doi.org/10.1016/j.physb.2011.12.021}
	{\path{doi:https://doi.org/10.1016/j.physb.2011.12.021}}.
	\newline\urlprefix\url{https://www.sciencedirect.com/science/article/pii/S0921452611012142}
	
	\bibitem{doi:10.1021/nn2007817}
	D.-K. Ko, C.~B. Murray, \href{https://doi.org/10.1021/nn2007817}{Probing the
		fermi energy level and the density of states distribution in pbte nanocrystal
		(quantum dot) solids by temperature-dependent thermopower measurements}, ACS
	Nano 5~(6) (2011) 4810--4817, pMID: 21506565.
	\newblock \href {http://arxiv.org/abs/https://doi.org/10.1021/nn2007817}
	{\path{arXiv:https://doi.org/10.1021/nn2007817}}, \href
	{https://doi.org/10.1021/nn2007817} {\path{doi:10.1021/nn2007817}}.
	\newline\urlprefix\url{https://doi.org/10.1021/nn2007817}
	
	\bibitem{Staring_1993}
	A.~A.~M. Staring, L.~W. Molenkamp, B.~W. Alphenaar, H.~van Houten, O.~J.~A.
	Buyk, M.~A.~A. Mabesoone, C.~W.~J. Beenakker, C.~T. Foxon,
	\href{https://doi.org/10.1209/0295-5075/22/1/011}{Coulomb-blockade
		oscillations in the thermopower of a quantum dot}, Europhysics Letters
	({EPL}) 22~(1) (1993) 57--62.
	\newblock \href {https://doi.org/10.1209/0295-5075/22/1/011}
	{\path{doi:10.1209/0295-5075/22/1/011}}.
	\newline\urlprefix\url{https://doi.org/10.1209/0295-5075/22/1/011}
	
	\bibitem{Zimbovskaya_2016}
	N.~A. Zimbovskaya,
	\href{https://doi.org/10.1088/0953-8984/28/18/183002}{Seebeck effect in
		molecular junctions}, Journal of Physics: Condensed Matter 28~(18) (2016)
	183002.
	\newblock \href {https://doi.org/10.1088/0953-8984/28/18/183002}
	{\path{doi:10.1088/0953-8984/28/18/183002}}.
	\newline\urlprefix\url{https://doi.org/10.1088/0953-8984/28/18/183002}
	
	\bibitem{PhysRevB.72.165308}
	A.~C. Johnson, J.~R. Petta, C.~M. Marcus, M.~P. Hanson, A.~C. Gossard,
	\href{https://link.aps.org/doi/10.1103/PhysRevB.72.165308}{Singlet-triplet
		spin blockade and charge sensing in a few-electron double quantum dot}, Phys.
	Rev. B 72 (2005) 165308.
	\newblock \href {https://doi.org/10.1103/PhysRevB.72.165308}
	{\path{doi:10.1103/PhysRevB.72.165308}}.
	\newline\urlprefix\url{https://link.aps.org/doi/10.1103/PhysRevB.72.165308}
	
	\bibitem{doi:10.1021/acs.nanolett.0c04017}
	S.~Dorsch, A.~Svilans, M.~Josefsson, B.~Goldozian, M.~Kumar, C.~Thelander,
	A.~Wacker, A.~Burke, \href{https://doi.org/10.1021/acs.nanolett.0c04017}{Heat
		driven transport in serial double quantum dot devices}, Nano Letters 21~(2)
	(2021) 988--994, pMID: 33459021.
	\newblock \href
	{http://arxiv.org/abs/https://doi.org/10.1021/acs.nanolett.0c04017}
	{\path{arXiv:https://doi.org/10.1021/acs.nanolett.0c04017}}, \href
	{https://doi.org/10.1021/acs.nanolett.0c04017}
	{\path{doi:10.1021/acs.nanolett.0c04017}}.
	\newline\urlprefix\url{https://doi.org/10.1021/acs.nanolett.0c04017}
	
	\bibitem{PhysRevB.72.205319}
	B.~Wunsch, M.~Braun, J.~K\"onig, D.~Pfannkuche,
	\href{https://link.aps.org/doi/10.1103/PhysRevB.72.205319}{Probing level
		renormalization by sequential transport through double quantum dots}, Phys.
	Rev. B 72 (2005) 205319.
	\newblock \href {https://doi.org/10.1103/PhysRevB.72.205319}
	{\path{doi:10.1103/PhysRevB.72.205319}}.
	\newline\urlprefix\url{https://link.aps.org/doi/10.1103/PhysRevB.72.205319}
	
	\bibitem{PhysRevB.80.165333}
	P.~Trocha, I.~Weymann, J.~Barna\ifmmode~\acute{s}\else \'{s}\fi{},
	\href{https://link.aps.org/doi/10.1103/PhysRevB.80.165333}{Negative tunnel
		magnetoresistance and differential conductance in transport through double
		quantum dots}, Phys. Rev. B 80 (2009) 165333.
	\newblock \href {https://doi.org/10.1103/PhysRevB.80.165333}
	{\path{doi:10.1103/PhysRevB.80.165333}}.
	\newline\urlprefix\url{https://link.aps.org/doi/10.1103/PhysRevB.80.165333}
	
	\bibitem{PhysicaE.64.254}
	N.~R. Abdullah, C.~S. Tang, A.~Manolescu, V.~Gudmundsson, {Delocalization of
		electrons by cavity photons in transport through a quantum dot molecule},
	Physica E 64 (2014) 254--262.
	
	\bibitem{PhysRevB.82.195325}
	N.~R. Abdullah, C.-S. Tang, V.~Gudmundsson,
	\href{http://link.aps.org/doi/10.1103/PhysRevB.82.195325}{{Time-dependent
			magnetotransport in an interacting double quantum wire with window
			coupling}}, Phys. Rev. B 82 (2010) 195325.
	\newblock \href {https://doi.org/10.1103/PhysRevB.82.195325}
	{\path{doi:10.1103/PhysRevB.82.195325}}.
	\newline\urlprefix\url{http://link.aps.org/doi/10.1103/PhysRevB.82.195325}
	
	\bibitem{PhysRevB.82.085311}
	V.~Moldoveanu, A.~Manolescu, V.~Gudmundsson,
	\href{http://link.aps.org/doi/10.1103/PhysRevB.82.085311}{{Dynamic
			correlations induced by Coulomb interactions in coupled quantum dots}}, Phys.
	Rev. B 82 (2010) 085311.
	\newblock \href {https://doi.org/10.1103/PhysRevB.82.085311}
	{\path{doi:10.1103/PhysRevB.82.085311}}.
	\newline\urlprefix\url{http://link.aps.org/doi/10.1103/PhysRevB.82.085311}
	
	\bibitem{Mao2016}
	J.~Mao, Z.~Liu, Z.~Ren,
	\href{https://doi.org/10.1038/npjquantmats.2016.28}{Size effect in
		thermoelectric materials}, npj Quantum Materials 1~(1) (2016) 16028.
	\newblock \href {https://doi.org/10.1038/npjquantmats.2016.28}
	{\path{doi:10.1038/npjquantmats.2016.28}}.
	\newline\urlprefix\url{https://doi.org/10.1038/npjquantmats.2016.28}
	
	\bibitem{doi:10.1063/1.3483618}
	S.~J. Shin, C.~S. Jung, B.~J. Park, T.~K. Yoon, J.~J. Lee, S.~J. Kim, J.~B.
	Choi, Y.~Takahashi, D.~G. Hasko,
	\href{https://doi.org/10.1063/1.3483618}{Si-based ultrasmall multiswitching
		single-electron transistor operating at room-temperature}, Applied Physics
	Letters 97~(10) (2010) 103101.
	\newblock \href {http://arxiv.org/abs/https://doi.org/10.1063/1.3483618}
	{\path{arXiv:https://doi.org/10.1063/1.3483618}}, \href
	{https://doi.org/10.1063/1.3483618} {\path{doi:10.1063/1.3483618}}.
	\newline\urlprefix\url{https://doi.org/10.1063/1.3483618}
	
	\bibitem{SVILANS20161096}
	A.~Svilans, M.~Leijnse, H.~Linke,
	\href{https://www.sciencedirect.com/science/article/pii/S1631070516300810}{Experiments
		on the thermoelectric properties of quantum dots}, Comptes Rendus Physique
	17~(10) (2016) 1096--1108, mesoscopic thermoelectric phenomena / Phénomènes
	thermoélectriques mésoscopiques.
	\newblock \href {https://doi.org/https://doi.org/10.1016/j.crhy.2016.08.002}
	{\path{doi:https://doi.org/10.1016/j.crhy.2016.08.002}}.
	\newline\urlprefix\url{https://www.sciencedirect.com/science/article/pii/S1631070516300810}
	
	\bibitem{PhysRevB.76.165432}
	P.~Trocha, J.~Barna\ifmmode~\acute{s}\else \'{s}\fi{},
	\href{https://link.aps.org/doi/10.1103/PhysRevB.76.165432}{Quantum
		interference and coulomb correlation effects in spin-polarized transport
		through two coupled quantum dots}, Phys. Rev. B 76 (2007) 165432.
	\newblock \href {https://doi.org/10.1103/PhysRevB.76.165432}
	{\path{doi:10.1103/PhysRevB.76.165432}}.
	\newline\urlprefix\url{https://link.aps.org/doi/10.1103/PhysRevB.76.165432}
	
	\bibitem{PhysRevB.70.081314}
	D.~Jacob, B.~Wunsch, D.~Pfannkuche,
	\href{https://link.aps.org/doi/10.1103/PhysRevB.70.081314}{Charge
		localization and isospin blockade in vertical double quantum dots}, Phys.
	Rev. B 70 (2004) 081314.
	\newblock \href {https://doi.org/10.1103/PhysRevB.70.081314}
	{\path{doi:10.1103/PhysRevB.70.081314}}.
	\newline\urlprefix\url{https://link.aps.org/doi/10.1103/PhysRevB.70.081314}
	
	\bibitem{PIROT2022413646}
	B.~R. Pirot, N.~R. Abdullah, A.~Manolescu, V.~Gudmundsson,
	\href{https://www.sciencedirect.com/science/article/pii/S0921452621007900}{Thermal
		transport controlled by intra- and inter-dot coulomb interactions in
		sequential and cotunneling serially-coupled double quantum dots}, Physica B:
	Condensed Matter 629 (2022) 413646.
	\newblock \href {https://doi.org/https://doi.org/10.1016/j.physb.2021.413646}
	{\path{doi:https://doi.org/10.1016/j.physb.2021.413646}}.
	\newline\urlprefix\url{https://www.sciencedirect.com/science/article/pii/S0921452621007900}
	
	\bibitem{AHMED2022413607}
	T.~Y. Ahmed, N.~R. Abdullah, V.~Gudmundsson,
	\href{https://www.sciencedirect.com/science/article/pii/S0921452621007560}{Controlling
		thermoelectric, heat, and energy currents through a quantum dot in sequential
		and cotunneling coulomb-blockade regimes}, Physica B: Condensed Matter 628
	(2022) 413607.
	\newblock \href {https://doi.org/https://doi.org/10.1016/j.physb.2021.413607}
	{\path{doi:https://doi.org/10.1016/j.physb.2021.413607}}.
	\newline\urlprefix\url{https://www.sciencedirect.com/science/article/pii/S0921452621007560}
	
	\bibitem{Goldozian2016}
	B.~Goldozian, F.~A. Damtie, G.~Kir{\v{s}}anskas, A.~Wacker,
	\href{https://doi.org/10.1038/srep22761}{Transport in serial spinful
		multiple-dot systems: The role of electron-electron interactions and
		coherences}, Scientific Reports 6~(1) (2016) 22761.
	\newblock \href {https://doi.org/10.1038/srep22761}
	{\path{doi:10.1038/srep22761}}.
	\newline\urlprefix\url{https://doi.org/10.1038/srep22761}
	
	\bibitem{Abdullah2017}
	N.~R. Abdullah, C.-S. Tang, A.~Manolescu, V.~Gudmundsson,
	\href{http://www.sciencedirect.com/science/article/pii/S1386947717311372}{{Spin-dependent
			heat and thermoelectric currents in a Rashba ring coupled to a photon
			cavity}}, Physica E: Low-dimensional Systems and Nanostructures (2017)
	--\href {https://doi.org/10.1016/j.physe.2017.09.011}
	{\path{doi:10.1016/j.physe.2017.09.011}}.
	\newline\urlprefix\url{http://www.sciencedirect.com/science/article/pii/S1386947717311372}
	
	\bibitem{KIRSANSKAS2017317}
	G.~Kiršanskas, J.~N. Pedersen, O.~Karlström, M.~Leijnse, A.~Wacker,
	\href{https://www.sciencedirect.com/science/article/pii/S0010465517302515}{Qmeq
		1.0: An open-source python package for calculations of transport through
		quantum dot devices}, Computer Physics Communications 221 (2017) 317--342.
	\newblock \href {https://doi.org/https://doi.org/10.1016/j.cpc.2017.07.024}
	{\path{doi:https://doi.org/10.1016/j.cpc.2017.07.024}}.
	\newline\urlprefix\url{https://www.sciencedirect.com/science/article/pii/S0010465517302515}
	
	\bibitem{GUDMUNDSSON20181672}
	V.~Gudmundsson, N.~R. Abdullah, A.~Sitek, H.-S. Goan, C.-S. Tang, A.~Manolescu,
	\href{http://www.sciencedirect.com/science/article/pii/S0375960118303748}{{Current
			correlations for the transport of interacting electrons through parallel
			quantum dots in a photon cavity}}, Physics Letters A 382~(25) (2018)
	1672--1678.
	\newblock \href {https://doi.org/10.1016/j.physleta.2018.04.017}
	{\path{doi:10.1016/j.physleta.2018.04.017}}.
	\newline\urlprefix\url{http://www.sciencedirect.com/science/article/pii/S0375960118303748}
	
	\bibitem{0953-8984-30-14-145303}
	N.~R. Abdullah, T.~Arnold, C.-S. Tang, A.~Manolescu, V.~Gudmundsson,
	\href{http://stacks.iop.org/0953-8984/30/i=14/a=145303}{{Photon-induced
			tunability of the thermospin current in a Rashba ring}}, Journal of Physics:
	Condensed Matter 30~(14) (2018) 145303.
	\newline\urlprefix\url{http://stacks.iop.org/0953-8984/30/i=14/a=145303}
	
	\bibitem{Goldhaber-Gordon1998}
	D.~Goldhaber-Gordon, H.~Shtrikman, D.~Mahalu, D.~Abusch-Magder, U.~Meirav,
	M.~A. Kastner, \href{https://doi.org/10.1038/34373}{Kondo effect in a
		single-electron transistor}, Nature 391~(6663) (1998) 156--159.
	\newblock \href {https://doi.org/10.1038/34373} {\path{doi:10.1038/34373}}.
	\newline\urlprefix\url{https://doi.org/10.1038/34373}
	
	\bibitem{Haake1973}
	F.~Haake, {Quantum Statistics in Optics and Solid-state Physics},
	\textit{Quantum Statistics in Optics and Solid-state Physics}, edited by G.
	Hohler and E.A. Niekisch, Springer Tracts in Modern Physics Vol. 66
	(Springer, Berlin, Heidelberg, New York, 1973, p. 98.).
	
	\bibitem{Breuer2002}
	H.-P. Breuer, F.~Petruccione, {The Theory of Open Quantum Systems}, Oxford
	University Press, Oxford, 2002.
	
	\bibitem{Nzar_2016_JPCM}
	N.~R. Abdullah, C.-S. Tang, A.~Manolescu, V.~Gudmundsson,
	\href{http://stacks.iop.org/0953-8984/28/i=37/a=375301}{{Competition of
			static magnetic and dynamic photon forces in electronic transport through a
			quantum dot}}, Journal of Physics: Condensed Matter 28~(37) (2016) 375301.
	\newline\urlprefix\url{http://stacks.iop.org/0953-8984/28/i=37/a=375301}
	
	\bibitem{ABDULLAH2018}
	N.~R. Abdullah,
	\href{http://www.sciencedirect.com/science/article/pii/S0375960118303177}{{Optical
			control of spin-dependent thermal transport in a quantum ring}}, Physics
	Letters A (2018).
	\newblock \href {https://doi.org/10.1016/j.physleta.2018.03.042}
	{\path{doi:10.1016/j.physleta.2018.03.042}}.
	\newline\urlprefix\url{http://www.sciencedirect.com/science/article/pii/S0375960118303177}
	
	\bibitem{PhysRevB.90.115313-2}
	M.~A. Sierra, D.~S{\'a}nchez,
	\href{https://link.aps.org/doi/10.1103/PhysRevB.90.115313}{Strongly nonlinear
		thermovoltage and heat dissipation in interacting quantum dots}, Phys. Rev. B
	90 (2014) 115313.
	\newblock \href {https://doi.org/10.1103/PhysRevB.90.115313}
	{\path{doi:10.1103/PhysRevB.90.115313}}.
	\newline\urlprefix\url{https://link.aps.org/doi/10.1103/PhysRevB.90.115313}
	
	\bibitem{Kouwenhoven_2001}
	L.~P. Kouwenhoven, D.~G. Austing, S.~Tarucha,
	\href{https://doi.org/10.1088%2F0034-4885%2F64%2F6%2F201}{Few-electron
		quantum dots}, Reports on Progress in Physics 64~(6) (2001) 701--736.
	\newblock \href {https://doi.org/10.1088/0034-4885/64/6/201}
	{\path{doi:10.1088/0034-4885/64/6/201}}.
	\newline\urlprefix\url{https://doi.org/10.1088%2F0034-4885%2F64%2F6%2F201}
	
	\bibitem{Nzar_ACS2016}
	N.~R. Abdullah, C.-S. Tang, A.~Manolescu, V.~Gudmundsson, ACS Photonics 3~(2)
	(2016) 249--254.
	\newblock \href
	{http://arxiv.org/abs/http://dx.doi.org/10.1021/acsphotonics.5b00532}
	{\path{arXiv:http://dx.doi.org/10.1021/acsphotonics.5b00532}}, \href
	{https://doi.org/10.1021/acsphotonics.5b00532}
	{\path{doi:10.1021/acsphotonics.5b00532}},
	\href{http://dx.doi.org/10.1021/acsphotonics.5b00532}{[link]}.
	\newline\urlprefix\url{http://dx.doi.org/10.1021/acsphotonics.5b00532}
	
	\bibitem{ABDULLAH2018199}
	N.~R. Abdullah, C.-S. Tang, A.~Manolescu, V.~Gudmundsson,
	\href{http://www.sciencedirect.com/science/article/pii/S0375960117311209}{{Effects
			of photon field on heat transport through a quantum wire attached to leads}},
	Physics Letters A 382~(4) (2018) 199--204.
	\newblock \href {https://doi.org/10.1016/j.physleta.2017.11.007}
	{\path{doi:10.1016/j.physleta.2017.11.007}}.
	\newline\urlprefix\url{http://www.sciencedirect.com/science/article/pii/S0375960117311209}
	
	\bibitem{Grabert1982}
	H.~Grabert, \href{https://doi.org/10.1007/BF01314753}{Nonlinear relaxation and
		fluctuations of damped quantum systems}, Zeitschrift f{\"u}r Physik B
	Condensed Matter 49~(2) (1982) 161--172.
	\newblock \href {https://doi.org/10.1007/BF01314753}
	{\path{doi:10.1007/BF01314753}}.
	\newline\urlprefix\url{https://doi.org/10.1007/BF01314753}
	
	\bibitem{PhysRev.89.728}
	R.~K. Wangsness, F.~Bloch,
	\href{https://link.aps.org/doi/10.1103/PhysRev.89.728}{The dynamical theory
		of nuclear induction}, Phys. Rev. 89 (1953) 728--739.
	\newblock \href {https://doi.org/10.1103/PhysRev.89.728}
	{\path{doi:10.1103/PhysRev.89.728}}.
	\newline\urlprefix\url{https://link.aps.org/doi/10.1103/PhysRev.89.728}
	
	\bibitem{PEDERSEN2010595}
	J.~N. Pedersen, A.~Wacker,
	\href{https://www.sciencedirect.com/science/article/pii/S1386947709002550}{Modeling
		of cotunneling in quantum dot systems}, Physica E: Low-dimensional Systems
	and Nanostructures 42~(3) (2010) 595--599, proceedings of the international
	conference Frontiers of Quantum and Mesoscopic Thermodynamics FQMT '08.
	\newblock \href {https://doi.org/https://doi.org/10.1016/j.physe.2009.06.069}
	{\path{doi:https://doi.org/10.1016/j.physe.2009.06.069}}.
	\newline\urlprefix\url{https://www.sciencedirect.com/science/article/pii/S1386947709002550}
	
	\bibitem{JONSSON201781}
	T.~H. Jonsson, A.~Manolescu, H.-S. Goan, N.~R. Abdullah, A.~Sitek, C.-S. Tang,
	V.~Gudmundsson,
	\href{http://www.sciencedirect.com/science/article/pii/S001046551730200X}{{Efficient
			determination of the Markovian time-evolution towards a steady-state of a
			complex open quantum system}}, Computer Physics Communications 220 (2017)
	81--90.
	\newblock \href {https://doi.org/10.1016/j.cpc.2017.06.018}
	{\path{doi:10.1016/j.cpc.2017.06.018}}.
	\newline\urlprefix\url{http://www.sciencedirect.com/science/article/pii/S001046551730200X}
	
\end{thebibliography}

\end{document}